\documentclass[reprint,amsmath,amssymb,aps,prb]{revtex4-1}
\usepackage{comment}
\usepackage{graphicx}
\usepackage{dcolumn}
\usepackage{bm}
\usepackage{hyperref}

\usepackage{amsmath,amsfonts}
\usepackage{color}
\newcolumntype{P}[1]{>{\centering\arraybackslash}p{#1}}

\begin{document}


\title{
Multistability and regime shifts in microbial communities explained by 
competition for essential nutrients}

\author{Veronika Dubinkina}
\thanks{These three authors contributed equally}
\affiliation{Department of Bioengineering and Carl R. Woese Institute for Genomic Biology,
University of Illinois at Urbana-Champaign, Urbana, IL 61801, USA}
\author{Yulia Fridman}
\thanks{These three authors contributed equally}
\affiliation{National Research Center "Kurchatov Institute", 
Akademika Kurchatova pl., Moscow, 123182, Russia}
\author{Parth Pratim Pandey}
\thanks{These three authors contributed equally}
\affiliation{Carl R. Woese Institute for Genomic Biology and 
National Center for Supercomputing Applications, 
University of Illinois at Urbana-Champaign, Urbana, IL 61801, USA}
\author{Sergei Maslov}
\thanks{maslov@illinois.edu}
\affiliation{
Department of Bioengineering and Carl R. Woese Institute for Genomic Biology,
University of Illinois at Urbana-Champaign, Urbana, IL 61801, USA.}

\date{\today}

\begin{abstract}
Microbial communities routinely have several possible species compositions or community states observed for the same environmental parameters. 
Changes in these parameters can trigger abrupt and persistent transitions (regime shifts) between such community states. 
Yet little is known about the main determinants and mechanisms of multistability in microbial communities. 
Here we introduce and study a resource-explicit model in which microbes compete for two types of essential nutrients. 
We adapt game-theoretical methods of the stable matching problem to identify all possible species compositions of a microbial community. 
We then classify them by their resilience against three types of perturbations: fluctuations in nutrient supply, invasions by new species, and small changes of abundances of existing ones. 
We observe multistability and explore an intricate network of regime shifts between stable states in our model. 
Our results suggest that multistability requires microbial species to have different stoichiometries of essential nutrients. 
We also find that balanced nutrient supply promote multistability and species diversity yet make individual community states less stable.
\end{abstract}

\maketitle
\section*{Introduction}
Recent metagenomics studies revealed that microbial communities collected in similar 
environments are often composed of rather different sets of species 
\cite{zhou2007differences,lahti2014tipping,Lozupone2012,Zhou2013,pagaling2017assembly,gonze2017multi}.
It remains unclear to what extent such alternative species 
compositions are deterministic 
as opposed to being an unpredictable outcome of communities' stochastic assembly. 
Furthermore, changes in environmental parameters may trigger abrupt and persistent transitions between these alternative species compositions \cite{Shade2012, rocha2018cascading, scheffer2003catastrophic}. Such transitions, known as ecosystem regime shifts, significantly alter the function of a microbial community and are difficult to reverse. Understanding mechanisms and principal determinants of alternative species compositions and shifts between them is practically important. Thus they have been extensively studied over the past several decades 
\cite{sutherland1974multiple,holling1973resilience,may1977thresholds,Tilman1997plantdiversity,Schroder2005,Fukami2011,bush2017oxicanoxic}.

Growth of microbial species is affected by many factors, 
with availability of nutrients being among the most important ones. 
Thus the supply of nutrients and competition for them
plays a crucial role in determining the species composition 
of a microbial community. The majority of modeling approaches 
explicitly taking nutrients into account are based on 
the classic MacArthur consumer-resource model and its variants 
\cite{macarthur1964competition,macarthur1970species,huisman2001biological,tikhonov2017collective,posfai2017metabolic,goldford2018emergent, goyal2018multiple,butler2018stability}.
This model assumes that every species co-utilizes several 
perfectly substitutable nutrients of a single type (e.g. 
carbon sources). However, it is well known that nutrients 
required for growth of a species exist in the form of several essential (non-substitutable) types including sources of C, N, P, Fe, etc. 
While real-life ecosystems driven by competition for multiple essential nutrients have been studied experimentally \cite{fanin2016synchronous, browning2017nutrient, camenzind2018nutrient}, the resource-explicit models capturing this type of growth are not so well developed beyond the foundational work by Tilman \cite{tilman1982resource}.

Here we introduce and study a new resource-explicit model of a microbial community 
supplied with multiple metabolites of two essential types.
%
%
This ecosystem is populated by microbes selected 
from a fixed pool of species.
%
%
%
We show that our model has a very large number of 
possible steady states classified by their species compositions. 
Using game-theoretical methods adapted from the well-known 
stable marriage problem \cite{gale1962college,gusfield1989stable}, 
we predict all of these states based only on the 
ranked lists of competitive abilities of individual 
species for each of the nutrients.
We further classify these states by their dynamic stability, 
and whether they could be invaded by other species in our pool.
We then focus our attention on a set of steady states that are both 
dynamically stable and resilient with respect to species invasion.

For each state we identify its feasibility 
range of all possible environmental parameters 
(nutrient supply rates) for which all 
of state's species are able to survive. 
We further demonstrate that for a given set of nutrient supply rates,
more than one state could be simultaneously feasible, thereby 
allowing for multistability. While the overall number of stable states in our model is exponentially large, 
only very few of them can be realized for a given set of environmental conditions quantified by nutrient supply rates. 
The principal component analysis of predicted microbial abundances in our model shows a separation between 
the alternative stable states reminiscent of real-life microbial ecosystems.
We further explore an intricate network of regime shifts between the alternative 
stable states in our model triggered by changes in nutrient supply. 
Our results suggest that multistability requires microbial species to have different stoichiometries of two essential resources. 
We also find that well-balanced nutrient supply rates matching the average species' stoichiometry promote multistability and species diversity yet make individual community states less structurally and dynamically stable. These and other insights from our resource-explicit model may help to understand the existing data and provide guidance for future experimental studies of alternative stable states and regime shifts in microbial communities.

\section{Results}
\subsection{Microbial community growing on two types of essential nutrients represented by multiple metabolites}
Our resource-explicit model describes a microbial ecosystem colonized by 
microbes selected from a pool of $S$ species.
Growth of each of these species could be limited by two types 
of essential resources, to which we refer to as ``carbon'' and ``nitrogen''. 
In principle, these could be any pair of resources essential for life: 
C, N, P, Fe, etc. 
A generalization of this model to more than two types of essential resources (e.g. C, N and P) is straightforward. 
Carbon and nitrogen resources exist 
in the environment in the form of $K$ distinct metabolites 
containing carbon 
, and $M$ other metabolites containing nitrogen.
For simplicity we ignore the possibility of the 
same metabolite providing both types. 
We further assume that each of the $S$ species in the pool is a 
specialist, capable of utilizing only a single pair of nutrients, 
i.e., one metabolite containing carbon and one metabolite containing nitrogen.

We assume that for 
given environmental concentrations of all nutrients, a growth rate of
a species $\alpha$ is limited by a single essential resource via 
Liebig's law of the minimum \cite{de1994liebig}:
\begin{equation}
g_{\alpha}(c,n)=\min(\lambda^{(c)}_{\alpha} c,\lambda^{(n)}_{\alpha} n) \quad .
\label{eq:growth_law}
\end{equation} 
Here $c$ and $n$ are the environmental concentrations of 
unique carbon and nitrogen resources consumed by this species.
The coefficients $\lambda^{(c)}_{\alpha}$ and $\lambda^{(n)}_{\alpha}$ are 
defined as species-specific growth rates per unit of concentration of each of two 
resources. They quantify the competitive abilities of the species $\alpha$ 
for its carbon and nitrogen resources, respectively. 
Indeed, according to the competitive exclusion principle, if two species are limited by 
the same resource, the one with the larger value of $\lambda$ wins the competition. 
Note that according to Liebig's law, if the 
carbon source is in a short supply so that $\lambda_{\alpha}^{(c)}c < \lambda_{\alpha}^{(n)}n$, 
it sets the value for this species growth rate. We refer to this situation as 
$c$-source {\it limiting} the growth of species $\alpha$. Conversely, when 
$\lambda^{(c)}_{\alpha}c > \lambda^{(n)}_{\alpha}n$, the $n$-source 
is {\it limiting} the growth of this species. Thus each species always
has exactly one growth-limiting resource and one non-limiting resource. 

\begin{figure*}
\centerline{\includegraphics[width=0.9\linewidth]{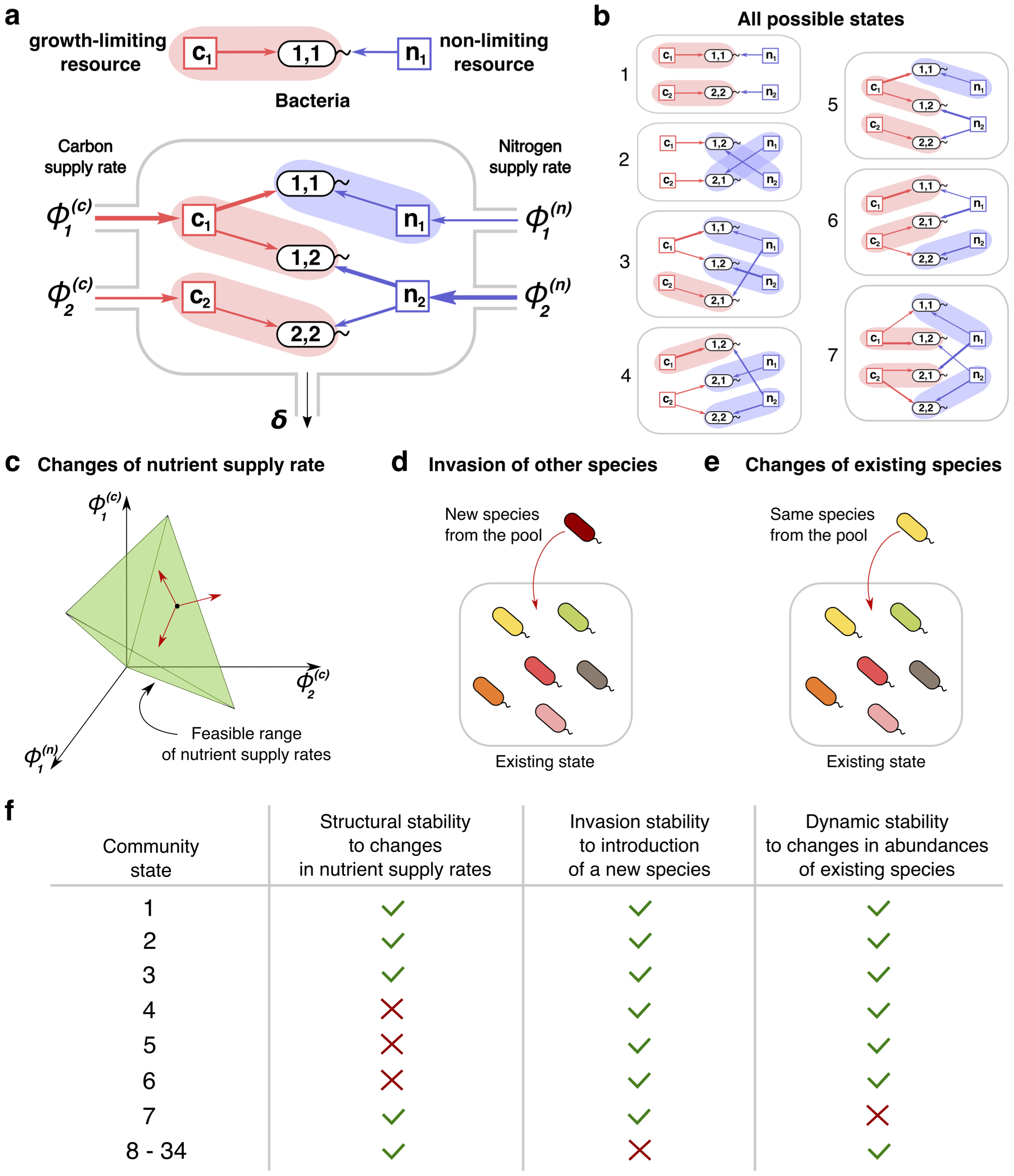}}
\caption{
{\bf 
Community states
and different types of their stability.} 
(\textbf{a}) A schematic depiction of the proposed experimental setup 
and one of several possible community states in the $2C\times2N\times4S$ 
model. Several sources of carbon an nitrogen are supplied at constant rates 
$\phi^{(c)}_i$ and $\phi^{(n)}_j$
to a chemostat with a dilution rate $\delta$. 
Red and blue square nodes represent these nutrients inside the chemostat with steady state concentrations $c_1$, $c_2$ (for carbon) and $n_1$, $n_2$ (for nitrogen). They are consumed by three microbial species labeled by the pair of carbon (the first index) and nitrogen (the second index) nutrients this species consumes. Shaded ovals connect every species to its unique growth-limiting nutrients. The fourth species $2, 1$ is not present in this steady state. (\textbf{b})  All 7 uninvadable states in the 
$2C\times2N\times4S$ model 
are depicted using the same schematic representation as in (a). 
Panels (\textbf{c}-\textbf{e}) schematically depict the three possible types of perturbations of a community state, corresponding to three different types of its stability. \textbf{(c)} Changes of nutrient supply rates, that may result in extinction of some of the species. Green shaded area schematically depicts the region of nutrient supply rates where a given state is feasible, red arrows represent the perturbations of nutrient supply rates. \textbf{(d)} Introduction of species currently absent from the system, i.e. invasion, that may change the set of surviving species. \textbf{(e)} Small fluctuations in abundances of existing species, that may disturb the dynamic equilibrium of the system and potentially drive it to another state. 
(\textbf{f}) Table that shows which stability criteria  are satisfied for $34$ possible states of the $2C\times2N\times4S$ model. Note that these types of stability are in general unrelated to each other.  
}
\label{fig1}
\end{figure*}

In our model microbes grow in a well-mixed chemostat-like 
environment subject to a constant dilution rate $\delta$ (see Fig. \ref{fig1}A 
for an illustration). The dynamics of the population density, $B_{\alpha}$, of 
a microbial species $\alpha$ is then governed by: 
\begin{equation}
\frac{dB_{\alpha}}{dt}=
B_{\alpha}\left[g_{\alpha}(c_i,n_j)-\delta \right] 
\quad , 
\label{eq:dbdt}
\end{equation}
where $c_i$ and $n_j$ are the specific pair of nutrients defining the growth rate 
$g_{\alpha}$ of this species according to the Liebig's law (Eq. \ref{eq:growth_law}). 
These nutrients are externally supplied 
at fixed rates $\phi_i^{(c)}$ and $\phi_j^{(n)}$ 
and their concentrations follow the equations:
\begin{eqnarray}
\frac{dc_i}{dt}&=&\phi_i^{(c)} -\delta \cdot c_i- 
\sum_{\text{all }\alpha \text{ using }c_i}
B_{\alpha}\frac{g_{\alpha}(c_i,n_j)}{Y_{\alpha}^{(c)}}\quad, \nonumber\\
\frac{dn_j}{dt}&=&\phi_j^{(n)} -\delta \cdot n_j- 
\sum_{\text{all }\alpha \text{ using }n_j}
B_{\alpha}\frac{g_{\alpha}(c_i,n_j)}{Y_{\alpha}^{(n)}} \quad .
\label{eq:dcdt}
\end{eqnarray}
Here $Y_{\alpha}^{(c)}$ and $Y_{\alpha}^{(n)}$ are the growth yields of 
the species $\alpha$ on its $c$- and $n$-resources respectively. Yields 
quantify the concentration 
of microbial cells generated per unit of concentration 
of each of these two consumed resources.  
The yield ratio $Y_{\alpha}^{(n)}/Y_{\alpha}^{(c)}$ determines 
the unique C:N stoichiometry of each species. 

A steady state of the microbial ecosystem can be found by setting the 
right hand sides of Eqs \ref{eq:dbdt}-\ref{eq:dcdt} to zero and solving them for 
environmental concentrations of all nutrients $c_i$, and $n_j$, 
and abundances $B_{\alpha}$ of all species. 
We choose to label all possible steady states by the list 
of species present in the state and {\it by the growth-limiting nutrient 
($c$ or $n$) for each of these species}. Thus, 
two identical sets of species, where at least one species 
is growth limited by a different 
nutrient are treated as two distinct states of our model. 
Conversely, our definition of a steady state
does not take into account species' abundances. Examples of such 
states in a system with 2 carbon, 2 nitrogen nutrients and 4 species 
(one species for every pair of carbon and nitrogen nutrients) with specific
values of species' competitive abilities $\lambda_{\alpha}^{(c)}$ and $\lambda_{\alpha}^{(n)}$ and yields 
$Y_{\alpha}^{(c)}$ and $Y_{\alpha}^{(n)}$ (see Supplementary Tables 1,2 for 
their exact values) are shown in Fig. \ref{fig1}B. For the sake of brevity we refer to 
this model as $2C\times2N\times4S$. 

Because each of the $S$ species in the pool could be absent from a given state,   
or, if present, could be limited by either its $c$- or its $n$-resource, 
the theoretical maximum of the number of distinct states is $3^S$ 
(equal to $81$ in our $2C\times2N\times4S$ example).
However, the actual number of possible steady states is considerably smaller
(equal to $34$ in this case).
Indeed, possible steady states in our model are 
constrained by a variant of the competitive exclusion principle 
\cite{gause1932experimental} (see Methods for details). 
One of the universal consequences of this principle is that the number of 
species present in a steady state of any consumer-resource model 
cannot exceed $K+M$ - the total number of nutrients. 
We greatly simplified the task of finding all steady 
states in our model by the discovery of the exact
correspondence between our system and a variant of the 
celebrated stable matching (or stable marriage) problem 
in game theory and economics \cite{gale1962college,gusfield1989stable}. 
(see Methods and Supplementary Materials, section \nameref{textsupp:st3}).


\subsection*{Three criteria for stability of microbial communities}

Each of the steady states identified in the previous chapter can be realized 
only for a certain range of nutrient supply rates. 
These ranges can be calculated using the steady state solutions 
of Eqs. \ref{eq:dbdt}, \ref{eq:dcdt}, governing the dynamics of 
microbial populations and nutrient concentrations respectively 
(see Methods). Among all formal mathematical solutions of these equations we 
select those where populations of all species 
and all nutrient concentrations are non-negative. 
This imposes constraints on nutrient supply rates, 
thereby determining their feasible range for a given steady state
(shown in green in Fig. \ref{fig1}C). 
The volume of such feasible range has been previously used to quantify 
the so-called structural stability of a steady state
\cite{rohr2014structural,grilli2017feasibility,butler2018stability}. 
States with larger feasible volumes generally tend to be more resilient 
with respect to fluctuations in nutrient supply. 

Stability of a community steady state could be also disturbed by a successful invasion of a 
new species (see Fig. \ref{fig1}D). We can test the resilience of a given state 
in our model with respect to invasions by other species.  
A state is called uninvadable if none of the other species 
from our pool can survive in the environment shaped by the existing species. 

In addition to structural and invasion types of stability 
described above, there is also a notion of dynamic stability 
of a steady state actively discussed in the ecosystems literature 
(see e.g. \cite{may1972will,allesina2012stability,butler2018stability}).
Dynamic stability can be tested by exposing a steady state to small perturbations 
in populations of all species present in this state (see Fig. \ref{fig1}E). 
The state is declared dynamically stable 
if after any such disturbance the system ultimately returns to its initial configuration 
(see Methods for details of the testing procedure used in our study).

We classify all of the steady states in our model according to these 
three types of stability.  
The example of this classification for the $2C\times2N\times4S$ model 
is summarized in Fig. \ref{fig1}F. Note, that in general, one type of 
stability does not imply another. Out of 34 possible steady states 
realized for different ranges of 
nutrient supply rates 
there are only 7 uninvadable ones. Unlike other consumer resource models, 
in our model the dynamic stability of a state with respect to species invasions 
does not depend on nutrient supply rates. 
In the $2C\times2N\times4S$ model only one of the states (labelled 7 in Fig. \ref{fig1}B) turned out to be dynamically unstable, while for the remaining 33 states small perturbations of microbial 
abundances present in the state did not trigger a change of the state. Unlike two other
types of stability, the structural stability has a continuous range. It could be 
quantified by the fraction of all possible combinations of nutrient supply rates 
for which a given state is feasible (referred to as state's normalized feasible range). 
We estimated normalized feasible ranges of all states in 
the $2C\times2N\times4S$ model 
using a Monte Carlo procedure described in Methods. The results are reflected in 
the second column of Fig. \ref{fig1}F, where 
a structurally stable state is defined as that whose 
normalized feasible range exceeds $0.1$.
In general we find that normalized 
feasible ranges of uninvadable 
states in our model have a broad log-normal distribution 
(see Fig. \ref{figs1} for details). 

It is natural to focus our attention on steady states that are simultaneously 
uninvadable and dynamically stable. Indeed, such states correspond to natural 
endpoints of the microbial community assembly process.  They would persist for as 
long as the nutrient supply rates do not change outside of their structural stability range. 
Therefore, they represent the states of microbial ecosystems that are likely to be experimentally observed. From now on we concentrate our study almost exclusively on those states and refer to them simply as stable states. 

\subsection{Regime shifts between alternative stable states}
%
\begin{figure*}
\centerline{\includegraphics[width=0.8\linewidth]{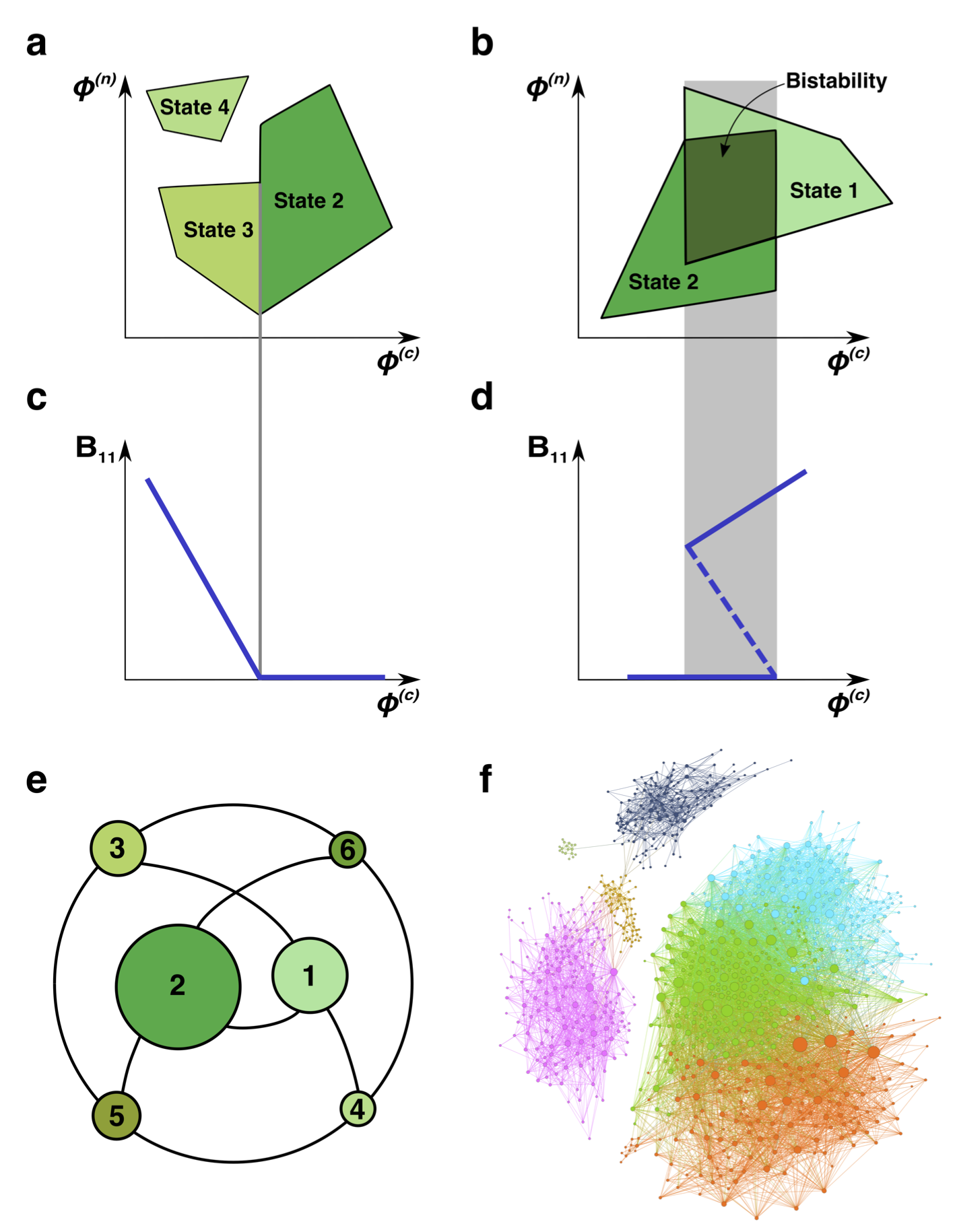}}
\caption{{\bf Regime shifts between alternative stable states.} (\textbf{a}) 
Shaded green areas schematically depict the feasible ranges of nutrient supply rates for several stable states in our model (\#2-\#4 in Fig. \ref{fig1}B). The feasible range of the state \#4 does not overlap with that of any other state. Feasible ranges of states \#2 and \#3 also do not overlap but share a common boundary.
Panel (\textbf{b}) depicts the opposite scenario of overlapping feasible ranges of another pair of stable states (\#1 and \#2 in Fig. \ref{fig1}B). In the overlapping region (dark green) they form a pair of alternative stable states. 
(\textbf{c}) A smooth transition between two states at the boundary. The population $B_{11}$ of the microbial species $(1,1)$ is plotted as a function of changing nutrient supply rate $\phi^{(c)}$ (same as the x-axis in panel (a)). Vertical gray line corresponds to the boundary between states \#3 and \#2. 
(\textbf{d}) A regime shift between two states. $B_{11}$ is plotted as a function of nutrient supply $\phi^{(c)}$ as it sweeps through the overlapping region (gray area) in panel (b). 
Note abrupt changes of $B_{11}$ at the boundaries of the overlapping region and its hysteretic behavior as expected for regime shifts. Dashed line corresponds to $B_{11}$ in a dynamically unstable state (\#7 in Fig. \ref{fig1}B). 
%
(\textbf{e}) The network of possible regime shifts between pairs of stable states in the $2C\times2N\times4S$ model. 
Each link represents a possible regime shift between two states it connects (overlap of their feasible ranges), nodes correspond to 6 uninvadable and dynamically stable states (state labels are the same as in Fig. \ref{fig1}B). Sizes of nodes reflect relative magnitudes of feasible ranges 
of states they represent. 
(\textbf{f}) Network of 8633 possible regime shifts between pairs 
of 893 uninvadable dynamically stable states in the $6C\times6N\times36S$ model. 
The size of each node reflects its degree (i.e. the total number of 
other stable states that a given state can shift into). The color of each node corresponds to its network modularity class calculated as described in Methods.
}
\label{fig2} 
\end{figure*}

The feasible ranges of nutrient supply of different stable states may or may not overlap with each other (see Fig. \ref{fig2}A-B for a schematic illustration of two different scenarios).
Whenever feasible ranges of two or more states overlap (see Fig. \ref{fig2}B) - multistability ensues. Note that the states in the overlapping region of their feasibility ranges 
constitute true alternative stable states defined and studied in the ecosystems literature \cite{sutherland1974multiple,holling1973resilience,may1977thresholds,Fukami2011,bush2017oxicanoxic}. The existence of alternative stable states goes hand-in-hand with regime shifts manifesting themselves as large discontinuous and hysteretic changes of species abundances \cite{scheffer2003catastrophic}. Every pair of states with overlapping feasibility ranges 
in our model corresponds to a possible regime shift between these states illustrated 
in Fig. \ref{fig2}D (note discontinuous changes in population $B_{11}$ of the microbial species $(1,1)$ at the boundaries of the overlapping region). Conversely, when feasible ranges of a pair of states do not overlap with each other but instead share a boundary (Fig. \ref{fig2}A), the transition between these states is smooth and non-hysteretic (Fig. \ref{fig2}C). It manifests itself in continuous changes in abundances of all microbial species at the boundary between states. 
%

As expected for regime shifts,
dynamically unstable states always accompany multistable regions in our model \cite{scheffer2003catastrophic} (see below for the detailed discussion of the interplay between multistability and dynamically unstable states). 
We observed that dynamically unstable state 7 in our $2C\times2N\times4S$ is 
feasible in the overlapping region between states 1 and 2 in Fig. \ref{fig2}B. The population $B_{11}$ in this state is shown as dashed line in Fig. \ref{fig2}D.       

We identified all possible regime shifts in the $2C\times2N\times4S$ model
by systematically looking for overlaps between feasible ranges of nutrient supply of all six uninvadable dynamically stable states. These regime shifts can be represented as a network in which nodes correspond to community's stable states and edges connect states with partially overlapping feasible ranges
(see Fig. \ref{fig2}E). 
Note that in general, this network does not capture overlaps between feasible ranges 
of more than two stable states, which will be discussed in the next section. Fig. \ref{fig2}F shows a much larger network of 8633 regime shifts
between 893 uninvadable dynamically stable states in the $6C\times6N\times36S$ variant of our model. In this model the microbial community is supplied with 6 carbon and 6 nitrogen nutrients and colonized from a pool of 36 microbial species (one for each pair of C and N nutrients) (see Supplementary Tables \ref{tab:Lambdas_c_L6}, \ref{tab:Lambdas_n_L6}, \ref{tab:Yields_c_L6}, \ref{tab:Yields_n_L6} for the values of 
$\lambda$'s and yields). 
For simplicity we did not show the remaining $165$ uninvadable stable states 
that have no possible regimes shifts to 
any other states. 
The size of a node is proportional to its degree (i.e. the total number of other states it 
overlaps with) ranging between 1 and 164 with average around 20 (degree distribution is shown 
in Fig. \ref{figs2}). 
%
%
The network modularity analysis (see Methods for details)
revealed 7 network modules indicating that pairs of states that could possibly undergo a 
regime shift are clustered together in the multi-dimensional space of nutrient supply rates.

\subsection{Patterns of multistability}
\begin{figure*}
\centerline{\includegraphics[width=\linewidth]{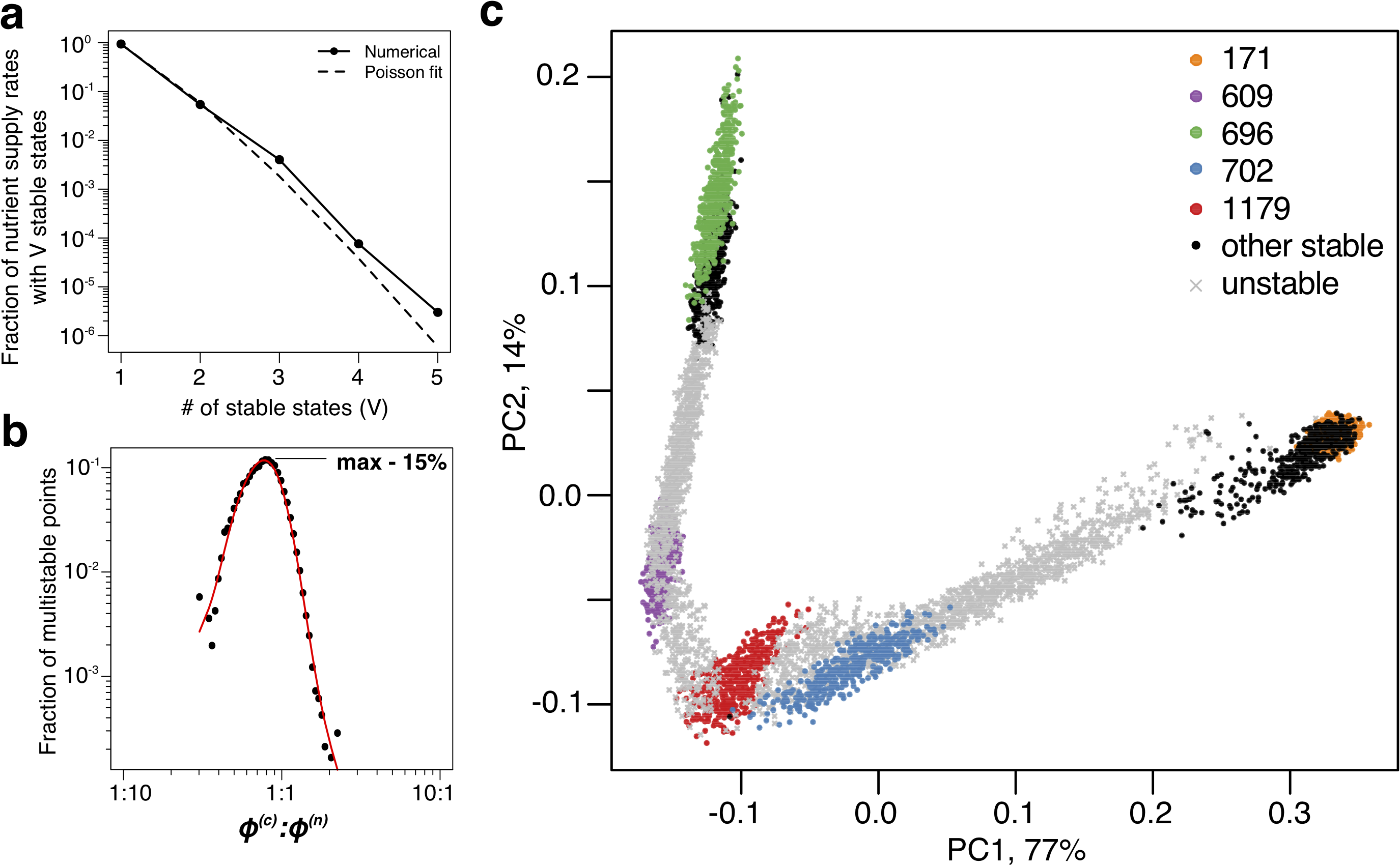}}
\caption{{\bf Patterns of multistability.} 
(\textbf{a}) The distribution of the number, $V$, of stable states
across the entire space of nutrient supply rates.  
The data is based on Monte Carlo sampling of 1 million different 
environments (combinations of nutrient supply rates) in the $6C\times6N\times36S$ model. 
Solid circles show the fraction of all sampled environments
for which $V=1,2,3,4,5$ stable states are simultaneously feasible. 
The dashed line is the fit to the data with a Poisson distribution for 
$v-1$ extra states giving rise to multistability.
(\textbf{b}) Fraction of multistable cases for different ratios of 
supply of two essential nutrients. The peak of the distribution 
is close to the balanced supply 
($\phi^{(c)}:\phi^{(n)}\simeq 1:1$). 
(\textbf{c}) The PCA plot of relative microbial abundances
in the vicinity of the environment, where $V=5$ stable states coexist. Supply rates were randomly sampled within $\pm$10\% from the initial 
environment. Each point shows 
the first (x-axis) and the second (y-axis) principal components 
of microbial abundances in every uninvadable state feasible for 
this combination of supply rates. Colored circles 
label the original five stable states, black circles - 
several other stable states, which became feasible for nearby supply rates, 
and grey crosses - dynamically unstable states feasible in this region of nutrient supply rates. 
}
\label{fig3}
\end{figure*}

In a general case, the number of stable states that are 
simultaneously feasible for given nutrient supply rates 
can be more than two. 
Furthermore, as the number of nutrients increases, 
the multistability with more than two stable states becomes 
progressively more common. 
In Fig. \ref{fig3}A we quantify the frequency 
with which $V$ multistable states occur 
in our $6C\times6N\times36S$ 
model across all possible nutrient supply rates
(see Methods for details of how this was estimated). 
$V-1$ approximately follows a Poisson distribution (dashed line in Fig. \ref{fig3}A) with $\lambda=0.063$.
Note that for some supply rates up to 5 stable states can be 
simultaneously feasible. However, the probability to find 
such cases is exponentially small. 

We further explored the factors that determine whether multistability is possible in resource-limited microbial communities and if yes, how common it is among different nutrient supply rates. 
Like in a simple special case of regime shift between two microbial species studied in Ref. \cite{tilman1982resource}, multistability in our model is only possible if individual microbial species have different C:N stoichiometry. This stoichiometry is given by the ratio 
of species' nitrogen and carbon yields. Our numerical simulations and mathematical arguments (see Supplementary Material, section \nameref{textsupp:st4}) 
show that when all species have exactly the same stoichiometry 
$Y^{(n)}_{\alpha}/Y^{(c)}_{\alpha}$, 
there is no multistability or dynamical instability in our model. 
That is to say, in this case for every set of nutrient supply rates the community has a unique uninvadable state, and all these states are dynamically stable. 
%
Simulations of the $2C\times2N\times4S$ example (see Fig. \ref{figs3}) show that the more similar is species' C:N stoichiometry 
(quantified by standard deviation of $\dfrac{Y_{ij}^{(n)}}{Y_{ij}^{(c)}}$), 
the less likely it is to find multistability among all possible 
nutrient supply rates (see Fig. \ref{figs3}). 


A complementary question is whether multistable states are more common around particular ratios of carbon and nitrogen supply rates. Fig. \ref{fig3}B shows this to be the case: the likelihood of multistability has a sharp peak around the well-balanced C:N nutrient supply rates. In this region multiple stable states are present for roughly 15\% of nutrient supply rate combinations. 
Note that the average C:N stoichiometry of species in our model is assumed to be 1:1. In a more general case, the peak of multistability is expected to be close to the average C:N stoichiometry of species in the community. 

To illustrate how multistable states manifest themselves in a commonly performed
Principle Component Analysis (PCA) of species' relative abundances,  
we picked the environment with  
$V=5$ simultaneously feasible stable states in our $6C\times6N\times36S$ model. 
In natural environments nutrient supply usually fluctuates both in time and space.
To simulate this we sampled a $\pm$10\% range of nutrient supply rates around this chosen environment (see Methods) and calculated species' relative abundances in each of the uninvadable states feasible for a given nutrient supply. To better understand the relationship between dynamically stable and unstable states we included 
the latter in our analysis. 
Fig. \ref{fig3}C shows the first vs the second principle components 
of relative microbial abundances sampled in this fluctuating environment.
(two more examples calculated for different multistable neighborhoods 
are shown in Fig. \ref{figs4}A-B).
One can see 5 distinct clusters each corresponding to 
a single dynamically stable uninvadable state. 
Interestingly, in the PCA plot these states are separated by $V-1=4$ dynamically unstable ones.
Furthermore, all states are aligned along a quasi-1D manifold with an alternating order of stable and unstable states. 
%

%
%




\subsection{Patterns of diversity and stability}
%
%
\begin{figure*}
\centerline{\includegraphics[width=\linewidth]{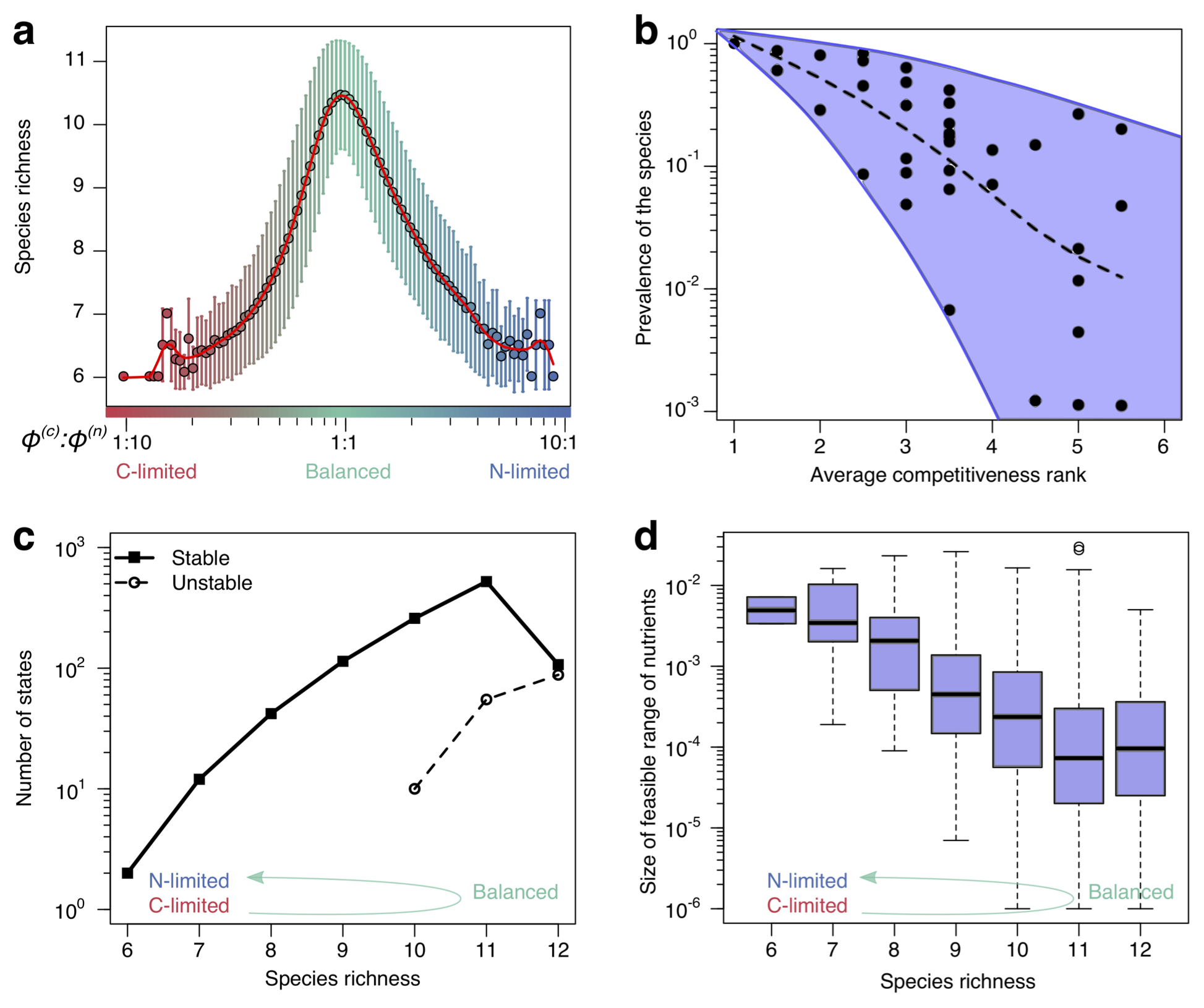}}
\caption{{\bf Patterns of diversity and stability
} 
(\textbf{a}) 
Average species richness (y-axis) of uninvadable stable states feasible 
for a given $\phi^{(c)}$: $\phi^{(n)}$ nutrient supply ratio (x-axis). 
Error bars correspond to standard deviation of species richness of individual states feasible for a given nutrient supply ratio.
(\textbf{b}): 
Scatter plot of the prevalence (y-axis) of each of the 36 species in the $6C\times6N\times36S$ model plotted vs its average competitiveness rank 
for its carbon and nitrogen sources. The latter is calculated from the rank 
order of $\lambda^{(c)}$ and $\lambda^{(n)}$ among all species consuming 
each resource (rank 1 corresponds to the largest $\lambda$ for 
this resource among all species). 
Species prevalence is quantified as the fraction of environments 
where a given species can survive.
The dashed line shows the average trend. 
(\textbf{c}) The number of uninvadable dynamically stable (solid line) and 
unstable (dashed line) states with a particular species richness (x-axis).
(\textbf{d})
Boxplot of nutrient feasibility ranges 
of uninvadable stable states plotted as a function of their species richness. 
All plots were calculated for the $6C\times6N\times36S$ model. 
}\label{fig4}
\end{figure*}

Above we demonstrated that multistable states are much more common for balanced nutrient supply rates, that is to say, when the average ratio of carbon and nitrogen supply rates matches the average C:N stoichiometry of species in the community (see Fig. \ref{fig3}B).
Interestingly, a balanced supply of nutrients also promotes species diversity. 
In Fig. \ref{fig4}A we plot the average number of species in a stable state, referred to as species richness, as a function of the average balance between carbon and nitrogen supplies
for $6C\times6N\times36S$ model. 
The species richness is the largest (around 10.5) for balanced nutrient supply rates, while  dropping down to the absolute minimal value of 6 in two extreme cases of very large imbalance of supply rates, where the nutrient supplied in excess becomes irrelevant in competition. 
In this case only 6 species that are top competitors for carbon metabolites (if nitrogen supply is plentiful) or, respectively nitrogen metabolites (if carbon is large) survive, while the rest of less competitive species are never present in uninvadable states. 

For balanced nutrient supply rates the relationship between species' competitiveness and its prevalence in the community is much less pronounced. 
It is shown in Fig. \ref{fig4}B where we plot the prevalence of the species as a function of its average competitiveness. 
%
Here the \textit{average competitiveness} of a species is defined as the mean of its rank to compete for its carbon and nitrogen resources. The rank $1$ being assigned to the most competitive species for a given resource (species with the largest value of $\lambda$), while the rank $6$ - to the least competitive species for this resource. Species \textit{prevalence} is given by the fraction of all environments where it can survive. 
%
Note that all $36$ species in our pool are present in some of the environments. 

In general more competitive species tend to survive in a larger subset of environments (see the dashed curve in Fig. \ref{fig4}B). 
For example, in our pool there is one species which happens to be the most competitive for both its carbon and nitrogen sources. This species is present in all of the states in every environment. 
%
However, we also find that some of the least competitive species (those at the right end of the x-axis in Fig. \ref{fig4}B) survive in a broad range of environments. For example, one species with average competitiveness rank of 5.5 corresponding to the last and next to last rank for its two resources still has relatively high 
prevalence of around $20\%$. 
This illustrates complex ways in which relative competitiveness of all species in the pool shapes their prevalence in a broad range of environments. 

We also explore the relationship between species richness of a state (i.e., its total number of surviving species) and its 
other properties. Fig. \ref{fig4}C shows an exponential increase of the number of 
uninvadable states as a function of species richness.
%
%
In our $6C\times6N\times36S$ model all uninvadable states with less than 10 species are dynamically stable (solid line in Fig. \ref{fig4}C), while those with 10 or more species can be both stable or unstable (dashed line in Fig. \ref{fig4}C). Overall the fraction of stable states to dynamically unstable ones decreases with species richness. In other words, the probability for a state to be dynamically unstable increases with the number of species. In this aspect our model behaves similar to the gLV model in Robert May's study \cite{may1972will}.

In Fig. \ref{fig4}D we show a negative correlation between the species richness of a stable state and its feasible range of nutrient supplies. Thus in our model the number of species in an ecosystem has a detrimental effect on the structural stability of the community quantifying its robustness to fluctuating nutrient supply \cite{rohr2014structural}. 
The empirically observed exponential decay of state's feasible range with its number of species is well described by a two-fold decrease per each species added (see Ref. \cite{servan2018coexistence} and \cite{grilli2017feasibility} for related results in the gLV model).
Note that the observed decrease in feasible range with species richness goes 
hand-in-hand with an increase in the overall number of states.  
Thus in well-balanced environments a large number of states are 
carving all possible combinations of nutrient supply into many small 
and overlapping ranges.  

Overall the results of our model with a large number of nutrients suggest the following picture. In nutrient-balanced environments
we expect to observe a high diversity of species in the existing communities. These species can form a very large number of possible combinations (uninvadable states). Each of these states could be realized only for a narrow range of nutrient supply rates indicating their low structural stability. Moreover in such environments we predict common appearance of multistability between some of these states.    

\section{Discussion}
The inspiration for our model was the common appearance of alternative stable states in ecosystems in general, and microbial communities in particular \cite{sutherland1974multiple,Tilman1997plantdiversity,Schroder2005,Fukami2011,bush2017oxicanoxic,pagaling2017assembly,gonze2017multi}. 
To the best of our knowledge our model is the first resource-explicit model capable of multistability between several states each characterized by a high 
diversity of species. We extend Tilman's scenario \cite{tilman1982resource} in which the growth of two species is limited by a pair of essential resources 
to the case of multiple nutrients of each type. This allows us to assemble complex communities with large number of co-existing species and provides additional insights into patterns of multistability in such communities. 

\subsection{Multistability requires diverse species stoichiometry and balanced nutrient supply} 
We find that multistability of microbial communities in our model 
requires species with different nutrient stoichiometries -- which is known to be highly variable in real microbes \cite{vrede2002elemental}.
In this aspect our model is similar to both the Tilman model 
\cite{tilman1982resource}, 
and the MacArthur family of models 
\cite{macarthur1964competition,macarthur1970species,chesson1990macarthur}. 
In common variants of the MacArthur model, the multistability is absent 
due to the assumption of identical nutrient yields of different species \cite{tikhonov2017collective,posfai2017metabolic,goldford2018emergent,goyal2018multiple,butler2018stability}. However, MacArthur model with different nutrient yields of different species should be capable of multistability. 
The larger is the variation of C:N stoichiometries of individual species in our model, the higher is the likelihood to observe multistability Fig. \ref{figs3}.
Somewhat unexpectedly, at least in the $2C\times2N\times4S$ model about half of the combinations of stoichiometries yielded no multistable states at all. Hence, variable stoichiometries do not guarantee multistability unless they are combined with the 
right combination of species competitiveness (see Section \nameref{textsupp:st4} in Supplementary Materials for mathematical arguments of why that may be the case).  

Another important factor favoring multistability in our model is the balanced supply of two essential nutrients (see Fig. \ref{fig3}B). 
It occurs when the average ratio of supply rates of two essential nutrients matches the average C:N stoichiometry of comminity's species (see Fig. \ref{fig3}B). When nutrient supplies are balanced, microbial community multistability is relatively common. Furthermore, for balanced nutrients the community can be in one of many different states, characterized by different combinations of limiting nutrients. These states tend to have high species diversity  (Fig. \ref{fig4}A) - a trend consistent with lake ecosystems in Ref. \cite{interlandi2001limiting}, and relatively small range of feasible supply rates (Fig. \ref{fig4}D). Hence, regime shifts can be easily triggered by changes in nutrient supply. 
The balanced region is characterized by a complex relationship between species competitiveness and survival, so that even relatively poor competitors could occasionally have high prevalence (species in the upper right corner of Fig. \ref{fig4}B).

In the opposite limit the supply of nutrients of one type (say nitrogen) greatly exceeds that of another type (say carbon). For such imbalanced supply the community has a unique uninvadable state, where every carbon nutrient supports the growth of the single most competitive species. Nitrogen nutrients are not limiting the growth of any species and thus have no impact on species survival and community diversity . 
As a consequence, the average diversity of microbial communities in such nutrient-imbalanced environments is low (about one half of that for balanced supply conditions). This is in agreement with many experimental studies showing that addition of high quantities of one essential nutrient (e.g. as nitrogen fertilizer) tends to decrease species diversity. This has been reported in numerous experimental studies cited in the chapter "Resource richness and species diversity" of Ref. \cite{tilman1982resource} as well as in recent experiments in microbial communities \cite{mello2016nutrient}.
\subsection{Interplay between diversity and stability in ecosystems with multiple essential nutrients}
Ever since Robert May's provocative question 
``Will a large complex system be stable?'' \cite{may1972will} 
the focus of many theoretical ecology studies has been on investigating the interplay between dynamic stability and species diversity in real and model ecosystems \cite{Ives2007}. 
May's prediction that ecosystems with large number of species tend to be 
dynamically unstable needs to be reconciled with the fact that we are 
surrounded by complex and diverse ecosystems that are apparently stable. 
Thus it is important to understand the factors affecting stability of ecosystems in general and microbial ecosystems in particular .

Here we explored the interplay between diversity and stability in a particular type of microbial ecosystems with multiple essential nutrients.
We discussed three criteria for stability of microbial communities shaped by the competition for nutrients: 
(i) how stable is the species composition of a community 
to fluctuations in nutrient supply rates; 
(ii) the extent of community's resilience to species invasions; 
and (iii) its dynamical stability to small 
stochastic changes in abundances of existing species. 
Naturally-occurring microbial communities may or may not be stable 
according to either one of these three criteria \cite{Ives2007}. 
The degree of importance of each single criterion is determined 
by multiple factors such as how constant are nutrient supply rates in time and space 
and frequently new microbial species migrate to 
the ecosystem. 

Our model provides the following insights 
into how these three criteria are connected to each other. 
First, as evident from Fig. \ref{fig1}F, the three types of 
stability are largely independent from each other. 
Second, communities growing on a well balanced mix of nutrients 
tend to have high species diversity (see peak in Fig. \ref{fig4}A). 
However, each of the community states in this regime 
tends to have a low structural stability with respect to nutrient fluctuations. 
In environments with highly variable nutrient supplies the community will 
frequently shift between these states. That is to say, some of the species 
will repeatedly go locally extinct and the vacated niches 
will be repopulated by others. Furthermore, many of the 
steady states in this regime are dynamically unstable giving rise 
to multistability and regime shifts.  In this sense
our model follows the general trend reported in 
Ref. \cite{may1972will}. Conversely, microbial communities growing on an imbalanced mix of essential 
nutrients have relatively low diversity (Fig. \ref{fig4}A) 
but are characterized by a 
high degree of structural and dynamic stability (see 
Fig. \ref{fig4}D and Fig. \ref{fig4}C respectively).

The existence of dynamically unstable states 
always goes hand in hand with multistability \cite{scheffer2003catastrophic}
(see Fig. \ref{fig2}B for an illustration of this effect in our model).
Interestingly, 
in our model we always find $V-1$ dynamically 
unstable states coexisting with $V$ dynamically 
stable ones for the same environmental parameters 
(see Fig. \ref{fig3}C and Fig. \ref{figs4} for some examples).
%
All states (both dynamically stable and unstable) shown in Fig. \ref{fig3}C 
are positioned along some one-dimensional curve in PCA coordinates.
This arrangement hints at the possibility of 
a non-convex one-dimensional Lyapunov function whose 
$V$ minima (corresponding to stable states) 
are always separated by $V-1$ maxima (unstable stable states) as 
dictated by the Morse theory \cite{milnor1963morse}. 
This should be contrasted with convex multi-dimensional Lyapunov 
functions used in Refs. \cite{macarthur1970species,case1979global,chesson1990macarthur}.

\subsection{Extensions of the model}
Our model can be extended to accommodate several 
additional properties of real-life microbial ecosystems:
First, one could include generalist species 
capable of using more than one nutrient of each type.
The growth rate of such species is given by: 
$$g_{\alpha}=
\min\left(\sum_{i \text{ used by }\alpha} \lambda^{(c)}_{\alpha i}c_i, 
\sum_{j \text{ used by }\alpha} \lambda^{(n)}_{\alpha j}n_j \; \right)$$
Here the sum over $i$ (respectively $j$) is carried out over all 
carbon (respectively nitrogen) sources that this species is capable of 
converting to its biomass. 
One may also consider the possibility of diauxic shifts between substitutable nutrient sources. In this case each generalist species is following a predetermined preference list of nutrients and uses its carbon and nitrogen resources one-at-a-time, as modelled in Ref. \cite{goyal2018multiple}. 
Since at any state each of the species is using a ``specialist strategy'', that is to say, it is growing on a single carbon and a single nitrogen source, we expect that many of the results of this study would be extendable to this model variant. Interestingly, the stable marriage problem can be used to predict the stable states of 
microbial communities with diauxic shifts between substitutable resources \cite{goyal2018multiple} and those in communities growing on a mix of two essential nutrients as in this study. It must be pointed out that these models use rather different variants of the stable marriage model.

It is straightforward to generalize our model to Monod's growth equation and to take into account non-zero death rate (or maintenance cost) of individual species (see Supplementary Materials section \nameref{textsupp:st1}).

One can extend our model to include 
cross-feeding between the species. In this case 
some of the nutrients are generated as 
metabolic byproducts by the species in the community.
These byproducts should be counted among nutrient sources and thus 
would allow the number of species to exceed the number of 
externally-supplied resources. 

Above we assumed a fixed size of the species pool. This constraint could be modified in favor of an expanding pool composed of a constantly growing number of species. These new species correspond to either migrants from outside of or mutate from outside of the community or mutants of the species within the community.  This variant of the model would allow one to explore the interplay between ecosystem's maturity (quantified by the number of species in the pool) and its properties such as multistability and propensity to regime shifts. 

\subsection{Control of microbial ecosystems exhibiting multistability and regime shifts}
In many practical situations we would like to be able to control microbial communities in a predictable and robust manner. That is to say, we would like to be able to reliably steer the community into one of its stable states and to maintain it there for as long as necessary. 
Alternative stable states and regimes shifts greatly complicate the task of manipulation and control of microbial ecosystems. Indeed, multistability means that the environmental parameters alone do not fully define the state of the community. In order to get it to a desired state, one needs to carefully select the trajectory along which one changes the environmental parameters (nutrient supply rates).
Changing these parameters could lead to disappearance (local extinction) of some microbial species
and open the ecosystem for colonization by others thereby changing its state.
Densely interconnected networks of regime shifts shown in Fig. \ref{fig2}E-F 
can be viewed as maps guiding the selection of the optimal trajectory to the desired stable species composition. 
The exploration of different manipulation strategies of microbial ecosystems is the subject of our future research \cite{maslov2019control}. 

\section{Acknowledgments}
Part of this work has been carried out at the University of Padova, Italy, in August 2018, during a scientific visit by one of us (S.M.). The authors wish to thank 
Prof. James O'Dwyer, Prof. Seppe Kuehn and Akshit Goyal for a critical reading of an earlier version of the manuscript. 

\section{Authors contributions}
S.M. designed the research; P.P. simulated the computational model; 
Y.F. and S.M. developed the theory and mapped this model to the stable marriage problem; 
P.P., S.M., and V.D. analyzed the data; S.M., V.D., and Y.F. wrote the manuscript; S.M. supervised the study.

\newpage
\section{Methods}
\subsection{Identification of all states and classification of them as invadable or uninvadable}

The competitive exclusion principle 
states that, in general, two species competing for the same growth-limiting nutrient cannot coexist with each other. 
Accounting for non-limiting nutrients present in our model, the competitive exclusion principle can be reformulated as the following two rules:

\begin{itemize}
\item Rule 1: In a given steady state each nutrient (either carbon or nitrogen) limits the growth of no more than one species. 
\item Rule 2: Any number of species can use a given nutrient in a non growth-limiting fashion. 
However, each of such species 
needs to be able to survive given 
the steady state concentration 
of this nutrient set by the growth-limited species.
That means that for every nutrient each of the non growth-limited species $\beta$ needs to be more competitive than the grow-limited species $\alpha$ for the same resource:
$\lambda^{(c)}_{\alpha} < \lambda^{(c)}_{\beta}$ (or $\lambda^{(n)}_{\alpha} < \lambda^{(n)}_{\beta}$ in case of a nitrogen nutrient).
\end{itemize}


Note that in any state of our model every species has a unique nutrient limiting its growth. By the virtue of the Rule 1, if a nutrient is limiting the growth of any species at all, such species is also unique. Hence, in a given state the relationship between surviving species and their growth-limiting nutrients (marked as shaded ovals in Fig. \ref{fig1}A) is an example of a matching on a graph of resource utilization. Rule 2 imposes additional limitations on this matching. 
As we show in the Supplementary Material, section \nameref{textsupp:st3}), uninvadable states correspond to stable matchings in a variant  of the celebrated stable marriage problem  \cite{gale1962college, gusfield1989stable}. 

Just like in the MacArthur
model \cite{macarthur1964competition} or any other resource-explicit model for that matter, the number of species present in a steady state of the community 
cannot exceed the total number of nutrients they 
consume. 
Any community constructed using Rules 1 and 2 represent a steady state of the ecosystem feasible for a certain range of nutrient supply rates.  This state can be either invadable or uninvadable, and either dynamically stable or not. 


%
For simplicity we work with an equal numbers of C and N resources ($L$ carbons and $L$ nitrogens), with one unique species capable of utilization of every pair of resources ($L^2$ species in total).  
We first selected the values of 
$\lambda^{(c)}_{(i,j)}$ and $\lambda^{(n)}_{(i,j)}$ 
from a uniform random distribution between 
10 and 100.
%
Note that all steady states of the community can be identified and tested for invadability using only the relative rank order of species' competitiveness for nutrients. For this we used the following exhaustive search algorithm:

\textit{Step 1} - Select
the subset of species whose growth is limited by C
(C-limited species). For every carbon nutrient there are $L$ ways to choose a C-limited species using this nutrient. Based on Rule 1 such C-limited species will be unique. There is also an additional possibility that this nutrient is not limiting growth of any species. The total number of possibilities is $L+1$ for each of $L$ carbon nutrients.
Thus, there are $(L+1)^L$ ways to choose the set of C-limited species and our algorithm will exhaustively investigate each of these possibilities one-by-one. 

\textit{Step 2} - Given the set of C-limited species selected in Step 1, we now select 
all N-limited species. 
We first eliminate from our search any species that doesn't have enough carbon to grow. That is to say, we go over all carbon nutrients one-by-one and eliminate all species whose $\lambda^{(c)}$ 
is smaller than that of the C-limited species (if any) for this carbon nutrient. 
Among the remaining species we go over the nitrogen nutrients
one-by-one and look for all possible ways to add a species limited by a given nitrogen source $n_j$ and satisfy the Rule 2. 
More specifically, we identify all species that use $n_j$ and can grow on their carbon sources (those species remained after the elimination procedure described above). We then compare $\lambda^{(n)}$s of these species to $\lambda^{(n)}$s of all C-limited species using $n_j$. To satisfy the Rule 2 for each $n_j$ we can add at most one N-limited species and its $\lambda^{(n)}$ has to be smaller than $\lambda^{(n)}$s of all C-limited species using $n_j$. Let $M_j$ be the number of such species ($M_j=0$ if there are no such species for a given $n_j$).
The total number of possible steady states of our model for a given combination of C-limited species selected in Step 1 is given by $\prod_{j=1}^{L}(M_j+1)$. Here the factor $M_j+1$ takes  into account an additional possibility to have no N-limited species for $n_j$. 

The unique way to construct an uninvadable state by following this algorithm is to go over all nitrogen sources one-by-one and for each of them attempt to add the N-limited species with the largest $\lambda^{(n)}$ among all species using this resource, whose growth is allowed by carbon constraints. If for every $n_j$ this species is allowed by the Rule 2, that is to say, if its $\lambda^{(n)}$ is smaller than $\lambda^{(n)}$ of all C-limited species using $n_j$, we successfully constructed a unique uninvadable state for a given set of C-limited species. Indeed, all possible invading species that are allowed to grow by their carbon nutrients will be blocked by their nitrogen nutrients. If, however, for any of $n_j$, the species with the largest $\lambda^{(n)}$ is not allowed by the Rule 2, that is to say, if its $\lambda^{(n)}$ is larger than $\lambda^{(n)}$ of at least one of the C-limited species, this species would make a successful invader of any state we construct. In this case there is no uninvadable state for the set of C-limited species selected during the Step 1. 

We used the above procedure to identify all possible steady states and to classify them as invadable and uninvadable for different numbers of resources used in our $2C\times2N\times4S$ and $6C\times6N\times36S$ examples. Note that, while this method 
is computationally possible for relatively small number of nutrients (we were able to successfully use it for up to 9 nutrients of each type), for larger systems one should rely on computationally more efficient algorithms based on the stable marriage problem \cite{gale1962college,gusfield1989stable}
as described in the Supplementary Material section \nameref{textsupp:st3}.

\bigbreak

\subsection{Monte-Carlo sampling of nutrient supply rates to identify feasible ranges of states}
\label{Meth:FeasibleVolumeAlgo}

Given the parameters defining all species (i.e., the set of their 
$\lambda$s and $Y$s) and the chemostat dilution constant $\delta$, each state $p$ 
is feasible within a finite region in the nutrient supply space (a $K+M$ dimensional space 
$\vec{\Phi}=\{\phi_i^{(c)}, \phi_j^{(n)}\}$)
where all microbial populations and nutrient concentrations are non-negative
and the limiting nutrients of every surviving species do not change. 
It is easy to show that in a steady state our system satisfies 
mass conservation laws for each of the nutrients:
\begin{eqnarray}
c_i+\sum_{\text{all }\alpha \text{ using }c_i}\frac{B_{\alpha}}{Y_{\alpha}^{(c)}}=\frac{\phi_i^{(c)}} {\delta} \nonumber \quad,\\
n_j+\sum_{\text{all }\alpha \text{ using }n_j}\frac{B_{\alpha}}{Y_{\alpha}^{(n)}}=\frac{\phi_i^{(n)}} {\delta} \quad.
\label{eq:conservation_laws}
\end{eqnarray}
%
%
To simplify the process of calculating the feasible volumes of all states we worked in the \textit{limit of high nutrient supply} where 
$\phi^{(c)}_{i} \gg \frac{\delta^2}{\lambda_{\alpha}^{(c)}}$ and 
$\phi^{(n)}_{j} \gg \frac{\delta^2}{\lambda_{\alpha}^{(n)}}$ for all species $\alpha$. 
In this case the concentration $\delta/\lambda^{(c,n)}_{\alpha}$ of any nutrient limiting growth of some species ($\alpha$ in this case) is negligible compared to its ``abiotic concentration''$\phi^{(c,n)}_{i}/\delta$, that is to say, its concentration 
before any microbial species were added to the chemostat. In this case one can ignore the terms $c_i$ and $n_j$ in 
Eqs. \ref{eq:conservation_laws} for all nutrient limiting growth of some species and leave only the 
ones that are not limiting the growth of any species. It is convenient to introduce the $K+M$-dimensional vector $\vec{X}_p$ 
of microbial abundances and non-limiting nutrient concentrations in a given state $p$. 
For example,  
for the uninvadable state \#5 in the $2C\times2N\times4S$ model we have: 
$\vec{X}_5 = \{B_{(1,1)}, B_{(1,2)}, B_{(2,2)}, n_2\}$.

The mass conservation laws (Eq. \ref{eq:conservation_laws}) can be used to obtain the feasible volumes of all states and can be represented in a compact matrix form for each state $p$: 
\begin{equation}
\vec{\Phi} = \hat{R_p} \vec{X}_p
\quad , 
\label{eq:Matrix-form}
\end{equation}
where $\Phi$ is the vector of $K+M$ nutrient supply rates and 
$\hat{R_p}$ is a matrix composed of inverse yields $Y^{-1}$ of surviving species and "1" for each of the non-limiting nutrients in a given state $p$. 
For example, for the state \#5 in our $2C\times2N\times4S$ model 
the Eq. \ref{eq:Matrix-form} expands to:

\begin{gather}
 \begin{bmatrix} \phi_1^{(c)} \\ \phi_2^{(c)} \\ \phi_1^{(n)} \\ \phi_2^{(n)} \end{bmatrix}
 =
  \begin{bmatrix}
   \frac{1}{Y_{(1,1)}^{(c)}} & \frac{1}{Y_{(1,2)}^{(c)}}  & 0 & 0 \\
   0 & 0 & \frac{1}{Y_{(2,2)}^{(c)}} & 0 \\
   \frac{1}{Y_{(1,1)}^{(n)}} & 0 & 0 & 0 \\
   0 & \frac{1}{Y_{(1,2)}^{(n)}} & \frac{1}{Y_{(2,2)}^{(n)}} & 1
   \end{bmatrix}
   \begin{bmatrix} B_{(1,1)} \\ B_{(1,2)} \\ B_{(2,2)} \\ n_2 \end{bmatrix}.
\label{eq:Matrix-form2}
\end{gather}

Using Eq. \ref{eq:Matrix-form} it is easy to check if a 
given state is feasible at a particular nutrient supply rate 
$\vec{\Phi}$ 
by multiplying $\hat{R_p}^{-1}$ (the inverse of the matrix $\hat{R_p}$) with 
$\vec{\Phi}$. If all of the elements of the resulting vector $\vec{X}_p$ are positive, 
then the 
state $p$ is feasible at $\vec{\Phi}$. If the 
matrix $\hat{R_p}$ 
is not invertible i.e., 
$\det(\hat{R_p})=0$, the state 
is feasible only on a low-dimensional 
subset of nutrient supply rates. 
This is not possible for a general choice of 
yields $Y$ and is not considered in our study.  

We imposed a common upper and lower bound on each of the $K+M$ 
nutrient supply rates \big( $\phi_{min} \leq \phi^{(c,n)}_i \leq \phi_{max} $\big) 
thus restricting the search of volumes of feasible states to a 
$(K+M)$-dimensional hypercube in the nutrient supply space. We chose 
$\phi_{min} = 10, \phi_{max}=1000$. The lower bound ensures that the
system is always in the limit of high nutrient supply since 
max($\frac{\delta^2}{\lambda_{\alpha}}$) $= 0.1 \ll  \phi_{min}=10$. 
We then randomly selected $10^6$ nutrient supply rate combinations $\vec{\Phi}$ within these bounds (Monte Carlo sampling) and checked the feasibility of each of the 33 possible states in the  $2C\times2N\times4S$ model 
and each of the 1211 uninvadable states in the  $6C\times6N\times36S$ model.
That is to say, for every set of nutrient supply rates $\vec{\Phi}$ and for every state $p$ we checked whether all elements of $\vec{X}_p$ are positive. 
The feasible range of nutrient supply rates of each state 
was estimated as the fraction of nutrient supply rate combinations (out of 1 million vectors $\vec{\Phi}$ sampled by our Monte Carlo algorithm) where it is feasible.


\bigbreak
\subsection{The network of regime shifts from  overlaps of feasible ranges} 
Two stable states are said to be capable of a regime shift if their feasibility ranges overlap with each other, i.e. if 
there exists at least one nutrient supply rate combination  
at which both these states are feasible. We used the data obtained by the Monte-Carlo sampling to look for such cases
and to construct 
networks shown in Fig. \ref{fig2}E, Fig. \ref{fig2}F. 
We used Gephi 0.9.2 software package to visualize the network in Fig. \ref{fig2}F and to perform its modularity analysis. Seven densely interconnected clusters shown with different colors in Fig. \ref{fig2}F were identified using Gephi's built-in module-detection algorithm \cite{blondel2008fast} with the resolution parameter set to 1.5. 


\bigbreak
\subsection{Dynamic stability of states}
\label{Meth:DynamicStabilityAlgo}
Each of the states in our model is either dynamically stable or dynamically unstable at all nutrient supply rates where it is feasible. We checked the dynamic stability of every 33 possible states (for the $2C\times2N\times4S$ model) and each of 1211 uninvadable states (for the $6C\times6N\times36S$ model) using the following two algorithms:

\begin{enumerate}
\item {\bf Small perturbation analysis} 
For $2C\times2N\times4S$ example we prepared each 
allowed state at multiple nutrient supply rate combinations where this state is feasible and subjected it to small perturbations of steady state values of all nutrient concentrations and of all  populations of species present in the state. 
We choose to perturb only the 
populations of species present in the state because an invadable state, by definition, would always be dynamically unstable against addition of a very small population of at least one invading species from the species pool. This instability should not render it dynamically unstable. 
The numerical integration of the system dynamics following 
a perturbation was done in C programming language 
using the CVODE solver library of the SUNDIALS 
package \cite{hindmarsh2005sundials}.



\item {\bf Inference of state's dynamic stability from the 
pattern of its overlaps with other states} 
The number of uninvadable states (1211) in our 
$6C\times6N\times36S$ model was too large to be tested directly as we did for the $2C\times2N\times4S$ model. 
Their dynamic stability was instead 
inferred from our Monte-Carlo simulations listing all feasible uninvadable states for every sampled nutrient supply rate combination.
We first identified 1022 uninvadable states which were the only feasible uninvadable state for at least one nutrient supply point.
All such states should be dynamically stable, since for every nutrient supply rate there should be at least one uninvadable dynamically stable state 
representing the end point of system's dynamics. 
The remaining 173 uninvadable states which were 
feasible for at least one of 1 million sampled nutrient supply rates were labelled as potentially dynamically unstable. Note that in our Monte-Carlo analysis we only sampled a finite (albeit large) number of supply rate combinations. 
Thus it is entirely possible that we missed some crucial 
supply rate combinations for which one of 
these states was the only uninvadable state. 
Any such point would have rendered this state as dynamically stable. Such false assignments might lead to a violation of the basic empirical rule in our model stating that $V$ uninvadable stable states are always accompanied by $V-1$ uninvadable dynamically unstable states ($V/(V-1)$ rule)
for some sampled nutrient supply rates. In our Monte Carlo simulations of the $6C\times6N\times36S$ model the $V/(V-1)$ rule was violated for only 370 nutrient supply rates combinations out of 1,000,000 sampled points. We believe that these violations were caused by an incorrect identification of dynamically unstable states mentioned above. To iteratively refine the lists of stable and unstable states, we went over all potentially unstable states one-by-one and checked whether reclassifying the state involved in the largest number of violations as stable would reduce the overall number of violations. If it did,  we reclassified this state as stable and recalculated the number of violations for all remaining points. By the end of this iterative procedure we were able to completely eliminate violations by reassigning 36 potentially unstable states as dynamically stable. This left us with $1022+36=1058$ dynamically stable and $173-36=137$ dynamically unstable uninvadable states in the $6C\times6N\times36S$ model. The remaining $1211-1058-137=16$ uninvadable states were not feasible for any of 1,000,000 sampled nutrient supply rates. Hence their dynamic stability remains unidentified. Both 36 reassigned states and 16 undetected states are expected to have very small ranges of feasible nutrient supply rates.
\end{enumerate}
\bigbreak

\subsection{Multistability as a function of variation in stoichiometric ratios of different species}

To investigate how the extent of multistability in our model depends on variation in stoichiometric ratios of different species, we simulated 4000 variants of the $2C\times2N\times4S$ model. In these variants we kept the same choice of species competitiveness (quantified by their $\lambda$s) but reassigned their yields $Y$. 
To cover a broad range of standard deviations of N:C stoichiometry of different species (their $Y_{\alpha}^{(c)}/Y_{\alpha}^{(n)}$) we randomly sampled yield combinations from gradually expanding intervals. First we simulated 1000 model variants, where yields of four species were independently drawn from $U(0.45, 0.55)$. These simulations were followed by 1000 model variants where yields of four species were drawn from $U(0.3, 0.7)$, 1000 model variants with yields from $U(0.1, 0.9)$ and ,finally, 1000 model variants with yields from $U(0.01, 1.0)$. In each variant of the model with a particular set of yields of 4 species we calculated the fraction of multistable points among $10^5$ nutrient supply rate combinations as described in the section \nameref{Meth:FeasibleVolumeAlgo} of Methods. 

The results are shown in Fig. \ref{figs3}. Its x-axis is the binned empirical standard deviation of species N:C stoichiometry equal to $Y^{(c)}/Y^{(n)}$, y-axis is the binned fraction of multistable nutrient supply rates, the color is proportional to $\log_{10}$ of the fraction of yield combinations (out of 4000 sampled yield combinations) that belong to a given bin of x- and y-axes.
\bigbreak

\subsection{GitHub repository of the code used in our project}

The PCA analysis, plots and statistical tests were implemented using R version 3.4.4. Other simulations were carried out in C (using compiler gcc version 5.4.0) and Python 3.5.2. Matlab analysis was done using MATLAB and Statistics Toolbox Release 2018a, The MathWorks, Inc., Natick, Massachusetts, United States. The code for both our simulations and statistical analysis can be downloaded from: 
\sloppy\url{https://github.com/ssm57/CandN}.


\bibliography{c_and_n_citations} 
\newpage

\renewcommand{\theequation}{S\arabic{equation}}
\setcounter{equation}{0}
\renewcommand{\thefigure}{S\arabic{figure}}
\setcounter{figure}{0}


\section{Supplementary Material}
\subsection{General form of growth laws} \label{textsupp:st1}
It is straightforward to generalize our model to allow a 
more general functional form for 
growth laws than Liebig's law, $\min(\lambda_{\alpha}^{(c)} c_i,
\lambda_{\alpha}^{(n)} n_j)$.  Microbial growth on 
two essential substrates is thought to normally 
follow Monod's equation for the rate-limiting nutrient:
$g^{(m)}_{\alpha}\min(c_i/(K_{\alpha}^{(c)}+c_i),n_j/(K_{\alpha}^{(n)}+n_j))$ 
(See Ref. \cite{kovrovkovar1998growth} for a discussion of 
limitations of Monod's law).
For low concentrations of the rate-limiting nutrient, say carbon source, 
the Monod's law simplifies to the linear growth law used throughout this study: 
$g_{\alpha}=\lambda^{(c)}_{\alpha} c_i$. Microbes' competitive abilities, also known as their 
specific affinities towards each substrate, are related to the parameters of 
Monod's law via 

\begin{equation}
\lambda_{\alpha}^{(c)}= \frac{g^{(m)}_{\alpha}}{K_{\alpha}^{(c)}}; \quad \lambda_{\alpha}^{(n)}= \frac{g^{(m)}_{\alpha}}{K_{\alpha}^{(n)}}
\end{equation}

In another variant of growth laws, two essential nutrients at low concentrations 
jointly affect the growth rate of the microbe: 
$g^{(m)}_{\alpha} c_i \cdot n_j/[(K_{\alpha}^{(c)}+c_i) \cdot (K_{\alpha}^{(n)}+n_j)]$ 
(see Ref. \cite{bader1978analysis} for a discussion of these and other forms of
double-substrate growth law).
For simplicity of mathematical calculation we limited 
this study to Liebig's law.  However, many of the essential results we obtained (e. g. 
possible multistability in a system where species have different yields) 
hold for any growth laws listed above.
In fact, the low concentration version of the previous growth law, where 
$g^{(m)}_{\alpha} c_i \cdot n_j$ has been studied by one of us in the context 
of autocatalytic growth of heteropolymers \cite{tkachenko2017onset}. 
Instead of exponentially replicating microbial species 
Ref. \cite{tkachenko2017onset} considers pairs of mutually catalytic (and thus
exponentially growing) complementary ``2-mers'' (a specific sequence of 
two consecutive monomers anywhere within a polymer chain). This minor 
difference complicates the math, while leaving the basic properties unchanged.
Just like in our system, where up to $2L$ species (out of $L^2$ candidates) 
may simultaneously survive in the steady state of an ecosystem grown on 
of $L$ carbon and $L$ nitrogen sources, the polymer systems have no more than 
$2Z$ 2-mer ``species'' (out of $Z^2$ candidates) 
surviving in the steady state with polymers having 
$Z$ possible monomers on their right ends and $Z$ possible monomers on their left 
ends. Many (but not all) results of this paper are largely consistent with the present study.
Note that for polymers the yields of all ``species'' are equal to 1, that is to say, 
one new 2-mer is formed upon ligation of one left end of a polymer 
with one right of another polymer chain. Yet, the model in Ref. \cite{tkachenko2017onset}
is capable of (at least) bistability. At present, it is not clear if this 
is due to autocatalytic cycles having length 2 or this property would survive
in a simpler version of the model in which instead of the Eq. (1) of Ref. \cite{tkachenko2017onset}
one has

$$
\dot{d}_{ij}=d_{ij}(\lambda_{ij} l_i r_j -\delta)
$$

and the overall fluxes of left and right ends are independent from each other 
(instead of both being equal to $c_i=\phi_i/\delta$ as in Ref. \cite{tkachenko2017onset}).

Another variant of the model is where each species $\alpha$ has its own unique ``death'' or ``maintenance'' rate $\delta_{\alpha}$, playing the role of the same dilution rate $\delta$. The steady states of this model (but not the dynamics leading to these states) can be calculated by 
dividing both sides of equations \ref{eq:dbdt} by 
$\delta_{\alpha}$. This is equivalent by redefining 
the competitiveness parameters to $\tilde{\lambda_{\alpha}}=\lambda_{\alpha}/\delta_{\alpha}$ and setting the chemostat dilution rate to 
$\tilde{\delta}=1$. All of our results in the high-flux regime $\phi \gg \delta^2/\lambda$ would remain unchanged.

From (Eq. \ref{eq:Matrix-form}-Eq. \ref{eq:Matrix-form2}) one can see that 
when all species have the same C:N stoichiometry, 
the maximal number of microbialspecies in a state is equal to the 
number of nutrients minus 1. Indeed, one can show that a state $p$ 
with $S_{surv}=K+M$ has $\det(\hat{R_p})=0$, which means that 
the feasible volume of any such state is zero. These states are 
only possible on a lower-dimensional manifold in the $(K+M)$-dimensional 
space of supply rates (these results have been already discussed by Tilman 
in his special case \cite{tilman1982resource}. 

Multistability is also possible in a variant of the MacArthur model
\cite{macarthur1964competition,macarthur1970species,chesson1990macarthur} in which different species have different yields on individual carbon sources \cite{maslov2018unpublished}. 
A convex Lyapunov function \cite{macarthur1970species} precluding 
multistability does not exist in this case. 

\subsection{Constraints on steady states from microbial and nutrient dynamics}\label{textsupp:st2}

A steady state of equations describing the 
microbial dynamics (Eq. \ref{eq:dbdt}) is realized
when either $B_{\alpha}=0$ (the species was absent from the system from the start 
or subsequently went extinct) or when its growth rate $g_{\alpha}$ is 
exactly equal to the chemostat dilution rate $\delta$. This imposes constraints 
on steady state nutrient concentrations with the number of constraints 
equal to the number of microbial species present with non-zero concentrations. 
Since, in general, the number of constraints cannot be larger than the 
number of constrained variables, no more than $K+M$ of species could be 
simultaneously present in a steady state of the ecosystem. For Liebig's 
growth law used in this study, each resource can have no more than 
one species for which this resource limits its growth, that is to say, 
which sets the value of the minimum in $\min(\lambda_{\alpha}^{(c)} c_i,
\lambda_{\alpha}^{(n)} n_j)$
The steady state concentrations of these resources are 
given by $c^{(*)}_i =\delta/\lambda_{\alpha}^{(c)}$ (if the growth is 
limited by the carbon source) 
and $n^{(*)}_j =\delta/\lambda_{\alpha}^{(n)}$ (if the growth is 
limited by the nitrogen source). Here $\alpha$ is the species 
whose growth is rate-limited by the 
resource in question. In a general case, no more than one species can be 
limited by the same resource (carbon in our example), since the species 
with the largest $\lambda^{(c)}$ would outcompete other 
species with smaller values of $\lambda^{(c)}$ by making the steady state concentration $c^{(*)}_i$ so low that other species can no longer grow on it. Note however, that multiple species $\beta$ could consume the same resource as the rate-limiting 
species $\alpha$, as long as their growth is not limited
by the resource. Each of these species must 
then be limited by 
their other nutrient (a nitrogen source in our example). However, 
their survival requires that carbon concentration set by species $\alpha$ is sufficient for their growth. Thereby, any species growing on a resource in a non-limited fashion must have $\lambda_{\beta}^{(c)} >\lambda_{\alpha}^{(c)}$.

Mathematically, it cane be proven by observing that, since species $\beta$ is limited by its nitrogen resource, one must have
$\lambda_{\beta}^{(c)} c^{(*)}_i >\lambda_{\beta}^{(n)} n^{(*)}_j$. 
At the same time in a steady state, the concentrations of all rate-limiting 
resources are determined by the dilution rate $\delta$ via 
$\lambda_{\beta}^{(n)} n^{(*)}_j=\delta$, and 
$\lambda_{\alpha}^{(c)} c^{(*)}_i=\delta$.  
Combining the above three expressions one gets:
$\lambda_{\beta}^{(c)} c^{(*)}_i >\lambda_{\beta}^{(n)} n^{(*)}_j=
\delta=\lambda_{\alpha}^{(c)} c^{(*)}_i$, or simply
$\lambda_{\beta}^{(c)} >\lambda_{\alpha}^{(c)}$.
The constraints on competitive abilities $\lambda$ 
for species present in a steady state in our model are then:
\begin{itemize}
\item Exclusion Rule 1: Each nutrient (either carbon or nitrogen source) 
can limit the growth of no more than one species $\alpha$.
From this it follows that the number of species co-existing in any given 
steady state cannot be larger than $K+M$, the total number of nutrients. 
\item Exclusion Rule 2: Each nutrient (e.g. specific carbon source) 
can be used by any number of species in a 
non-rate-limiting fashion (that is to say, where it does not constrain 
species growth in Liebig's law). However, any such species $\beta$ 
has to have $\lambda_{\beta}^{(c)} >\lambda_{\alpha}^{(c)}$, 
where $\lambda_{\alpha}^{(c)}$ is the competitive ability of 
the species whose growth is limited by this nutrient. 
In case of a nitrogen nutrient, the constraint becomes 
$\lambda_{\beta}^{(n)} >\lambda_{\alpha}^{(n)}$. 
\end{itemize}

Note that the steady state solutions of equations 
Eq. \ref{eq:dbdt} do not depend on populations 
$B_{\alpha}$ of surviving species. Their steady state populations
$B^{(*)}_{\alpha}$ are instead determined by Eq. \ref{eq:dcdt}. 
Taking into account that, in a steady state, 
the growth rate of each surviving  species is exactly 
equal to the dilution rate $\delta$ of the chemostat, 
after simplifications one gets:  
\begin{eqnarray}
\frac{\phi_j^{(c)}}{\delta} &=&c^{(*)}_i+
\sum_{\text{all }\alpha \text{ using }c_i}
\frac{B^{(*)}_{\alpha}}{Y_{\alpha}^{(c)}}
\nonumber \\
\frac{\phi_j^{(n)}}{\delta}&=&n^{(*)}_j+ 
\sum_{\text{all }\alpha \text{ using }n_j}
\frac{B^{(*)}_{\alpha}}{Y_{\alpha}^{(n)}}
\label{eq:ss_nc}
\end{eqnarray}
As described above, the steady state 
concentration of resources 
are given by 
$\delta/\lambda^{(\text{c or n})}_{\alpha}$, where 
$\alpha$ are the species rate-limited 
by each resource. In the absence of such species, the 
concentration of a resource is given by anything left 
after it being consumed by surviving species in a non-rate-limiting 
manner. One can show that in this case, 
the resource (e.g. carbon) concentration has to be 
larger than $\delta/\lambda_{\beta}^{(c)}$, where 
$\lambda_{\beta}^{(c)}$ is the smallest affinity 
among microbes utilizing this resource.

One convenient approximation greatly simplifying working with Eq. \ref{eq:ss_nc} is the ``high-flux limit'' in 
which $\phi^{(c)}_i \gg \delta^2/\lambda_{\alpha}^{(c)}$ and 
$\phi^{(n)}_j \gg \delta^2/\lambda_{\alpha}^{(n)}$. In this 
approximation one can approximately set to zero the steady state concentrations of 
all resources that have a species rate-limited by them. The steady state 
concentrations of the remaining resources can take any value as long 
as it is positive. Hence, in this limit the 
Eq. \ref{eq:ss_nc} can be viewed as a simple matrix 
test of whether a given set of surviving species limited by a 
given set of resources is possible for a given set of nutrient fluxes. 
Indeed, my multiplying the vector of fluxes with the 
inverse of the matrix $\hat{R}$ composed of inverse yields of surviving
species and 1 for nutrients not limiting the growth of any species one formally
gets the only possible set of steady state species abundances, $B^{(*)}_{\alpha}$, 
and a subset of non-limiting resource concentrations $c^{(*)}_i$ and $n^{(*)}_j$. 
If all of them are strictly positive - the steady state is possible. 
If just one of them enters the negative territory - the steady state 
cannot be realized for these fluxes of nutrients. 

The above rule can be modified to apply even below the 
high-flux limit with the following modifications:
1) Instead of $\phi^{(c)}$ (or $\phi^{(n)}$), one uses 
their ``effective values'' $\tilde{\phi}^{(c)}$ (or $\tilde{\phi}^{(n)}$) introduced in \cite{goyal2018multiple}, determined as 
\begin{eqnarray}
\tilde{\phi}^{(c)}_i&=&\phi^{(c)}_i-\frac{\delta^2}{\lambda^{(c)}_{\alpha(i)}} \nonumber \\
\tilde{\phi}^{(n)}_j&=&\phi^{(n)}_j-\frac{\delta^2}{\lambda^{(n)}_{\alpha(j)}} \quad ,
\label{eq:offset}
\end{eqnarray}
where $\alpha(i)$ is the (unique) species limited by the 
nutrient $i$. If the nutrient is not limiting for any os the  species in the steady state, 
$\alpha(i)$ is the species using the nutrient in a non-limited fashion, which has the {\it smallest} value of $\lambda$. This last rule comes from the observation that in order for a non-limiting resource not to become limiting for a species $\beta$ currently using it in a non-limiting fashion, its concentration cannot fall below 
$\delta/\lambda^{(x)}_{\beta}$. Thus, when checking the feasibility of a given state, the concentration of a non-limiting resource can be written as $\delta/\lambda^{(x)}_{\beta}+$ a positive number, or (more conveniently) the influx of this resource can be offset as described in Eqs. \ref{eq:offset}



\subsection{Stable matching approach for 
identification and classification of steady 
states}\label{textsupp:st3}

First we describe the exact one-to-one mapping between 
all uninvadable steady states (UIS) 
in our model and the complete set of ``stable marriages''
in a variant of a well-known stable marriage or 
stable allocation problem developed by Gale and 
Shapley in the 1960s 
\cite{gale1962college} and awarded the Nobel 
prize in economics in 2012. This mapping provides 
us with constructive algorithms to identify and 
count all uninvadable steady states in our ecosystem.

We start by considering a special case of 
our problem with $L$ carbon and $L$ 
nitrogen sources and a pool of $L^2$ species,  
such that for every pair of sources $c_i$ (carbon) 
and $n_j$ (nitrogen) there is exactly one microbe 
$B_{ij}$ capable of using them.  For the sake of 
simplicity we have switched the notation 
from $B_{\alpha}$ to $B_{ij}$, where $\alpha=(ij)$
is the unique microbe in our pool capable of growing 
on $c_i$ and $n_j$.
Having considered this simpler situation we will return 
to the most general case of unequal numbers 
of carbon ($K$) and nitrogen ($M$) resources and any 
number of microbes from a pool of $S$ species competing 
for a given pair of resources.

This is where we need to revise in certain ways the
network representation of a steady state used in the main 
text (see Fig. \ref{fig1}A).  In the marriage game related theory,
the notion of (stable or unstable) matching explicitly 
refers to a bipartite graph with two distinct sets of vertices
and edges arranged in such a way that each one may join only 
a pair of elements belonging to different sets.  In our case,
it is natural to consider two sets of resource nodes (vertices),
one including all carbon nodes and the other one containing 
all nitrogen nodes.  An edge, or link, will appear between 
a carbon $c_i$ and nitrogen $n_j$ node if the microbe $B_{ij}$ using 
these two nutrients is present in the state represented by this 
particular bipartite network.

Furthermore, the specifics of our version of "marriage game",
or rather "residents vs hospitals", problem requires us
to consider {\itshape directed} bipartite graphs as the steady
state representations in our model.  For any species
$B_{ij}$ present in a given state, we choose the direction
of the edge joining node $c_i$ with node $n_j$ to be  
pointing from $c_i$ to $n_j$ if the microbe is limited by its
carbon nutrient (and the other way around, from $n_j$ to $c_i$, 
on the case of $B_{ij}$ being nitrogen-limited).
Fig. \ref{fig5} A shows the directed
bipartite graph representation of the state \#5
of a particular example of $2C\times2N\times4S$ system 
considered in the "Results" section of the main text.

In what follows we will refer to a resource as 
{\itshape occupied} if in a given steady state 
there is a microbe for
which this resource is rate-limiting.  In our 
bipartite network representation occupied resources have 
an outgoing edge (their out-degree is equal to 1), while 
unoccupied resources  have out-degree equal to 0.  
%

\subsubsection*{Review of results about stable matchings 
in the hospitals/residents problem}
The hospitals/residents problem \cite{gale1962college} 
is known in various settings.  The one directly relevant 
to our problem
is the following.  There are $L$ applicants for residency positions in 
$H \le L$ hospitals.  A hospital number $i$ has
$V_{i}$ vacancies for residents to fill, $V_{i}$
ranging from zero to $L$, $\sum V_i = L$.  Each hospital has a list
of preferences in which residency applicants are strictly
ordered by their ranks, from $1$ (the most desirable) to $L$ ,
(the least desirable).  These lists are generally different for different 
hospitals. Each applicant has a ranked list of preferred hospitals 
ranging from $1$ (the most desirable) to $H$ (the least desirable).
Those lists can also vary between applicants.
A {\itshape matching} is an assignment of applicants to hospitals 
such that all applicants got residency and all hospital vacancies are filled. 
A matching is {\itshape unstable}
if there is at least one applicant $a$
and hospital $h$ to which $a$ is not assigned 
such that:
\begin{enumerate}
\item Condition 1. Applicant $a$ prefers hospital $h$ to his/her assigned hospital;
\item Condition 2. Hospital $h$ prefers applicant $a$ to at least one of its assigned applicants.
\end{enumerate}
If such a pair $(a, h)$ exists, it is called ``a blocking
pair'' or ``a pair that blocks the matching''.
A {\itshape stable} matching by definition has
no blocking pairs.  
Gale and Shapley proved that for any set of applicant/hospital 
rankings and hospital vacancies there is at least one 
stable matching \cite{gale1962college}.
Generally the number of stable matching is larger than 
one. For example, for stable marriages and random rankings
the average number of stable matchings is given by 
$L/e \log L$ \cite{gusfield1989stable}. To the best 
of our knowledge, the dependence of this number on 
the distribution of hospital vacancies has not been 
investigated. 
The fact that the actual number of 
uninvadable states is rather close to its lower bound 
(compare black symbols and dashed line in Fig. \ref{fig1})
indicates that, at least for $L \leq 9$, the number 
of stable matchings averaged over all possible 
in-degree allocations is rather close to 1.

Gale and Shapely not only proved the existence of at least 
one stable matching, but also proposed a constructive algorithm 
on how to find it. Listed below are the main steps in this algorithm
optimized for  for applicants.
Each applicant first submits his/her application to the hospital 
ranking $1$ in his/her preference lists. Each hospital considers all 
applications it received so far and accepts all of the applicants 
if their number is less or equal than hospital's announced number of vacancies, 
$L_i$. If the number of applicants exceeds $L_i$, the hospital
gives a conditional admission to the
best-ranking $L_i$ applicants according to hospital's 
own preference list. Each applicant not admitted to their top hospital 
goes a step down on his/her preference
list and applies to the second-best hospital.  The latter admits
this applicant if (1) this hospital has not yet filled all of its vacancies 
or (2) all vacancies are filled, but among the conditionally admitted 
applicants there is at least one who ranks lower (according to hospital's list)
than the new applicant.  Such lower-ranked applicants 
are declined admission and replaced with better ones. They subsequently 
lower their expectations and apply to the next hospital on their list.  
After a number of iterations all applicants are admitted and 
all vacancies are filled so that this process stops.
As Gale and Shapley proved in Ref. \cite{gale1962college}, 
the resulting matching is stable. Furthermore, the theorem 
states that in this matching every applicant gets admitted 
to the best hospital among all stable matchings, while every hospital 
gets the worst set of residents among all stable matchings.
Later research described in Ref. \cite{gusfield1989stable}
describe more complex constructive algorithms allowing one to efficiently 
find all of the stable matchings starting with the applicant-optimal one.

Well developed mathematical apparatus of stable matching problem 
provides an invaluable help in the task of identifying all uninvadable
states in microbial ecosystems. Indeed, without its assistance 
this task would require exponentially long time.
To connect the problem of finding all uninvadable states to that of
finding all stable matchings between hospitals and residents, 
we start with the following three observations:

1) In any uninvadable steady state, either all carbon
sources or all nitrogen sources (or both) are occupied.  Indeed, if
in a steady state a carbon source $c_i$ and a nitrogen source $n_j$ are not-limiting 
to any microbes, then microbe $B_{ij}$ can always grow and thereby invade 
this state.  Thus uninvadable states can be counted separately:
one first counts the states where all nitrogen sources are occupied, and 
then counts those in which all carbon sources are occupied.  
Double counting happens when both carbon 
and all nitrogen sources are occupied.  We will keep the possibility 
of double counting in mind and return to this problem later.

2) For a pool of species, where for every pair of resources 
there is exactly one microbe using each (carbon, nitrogen) pair,
one can think of each of $L$ carbon (alternatively, nitrogen) 
sources as if it had a list of ``preferences''
ranking all nitrogen (correspondingly carbon) sources. 
Indeed, the ranking of competitive abilities $\lambda^{(c)}_{ik}$ 
of different microbes using the same carbon source $c_i$ but 
different nitrogen sources $n_k$ can be viewed as the ranking of 
nitrogen sources $k$ by the carbon source $i$.
Conversely, the ranking of $\lambda^{(n)}_{mj}$ 
with the same $n_j$ but variable 
$c_m$ can be thought of as ranking of 
carbon sources $c_m$ by the nitrogen source $n_j$.

\begin{figure*}
\centerline{\includegraphics[width=0.9\linewidth]{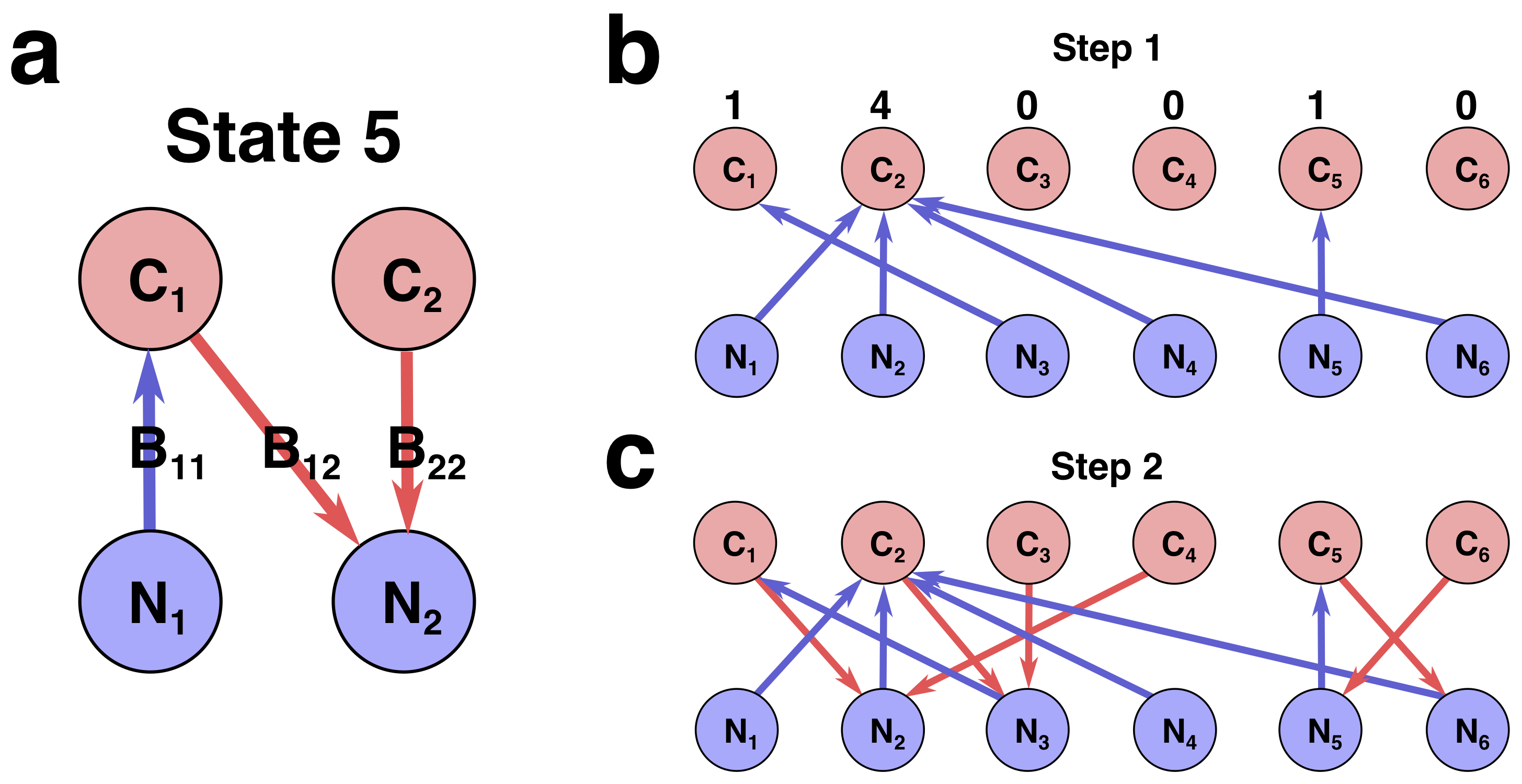}}
\caption{{\bf Network representation of a state in Stable Marriage analogy.} 
(\textbf{a}) Schematic representation of state \#5 in the $2C\times2N\times4S$ model (the same state is shown in Fig. \ref{fig1}A in Main text). Here each species represented as an arrow connecting two resources it is utilizing for growth, with the color and direction of arrow representing growth-limitation of a species (red corresponds to C-limited species, blue N-limited one). 
(\textbf{b})-(\textbf{c}) Schematic representation of two-step construction of state \#991 in the $6C\times6N\times36S$ model. 
(\textbf{b}) We first assign all species that are growth-limited by N (blue links outgoing from N sources). The numbers above C sources indicate number of vacancies for a given resource.
(\textbf{c}) Then for a given set of N-limited species we populate the remaining C-limited ones that are allowed by the Condition 2 (red links outgoing from C sources). 
}
\label{fig5}
\end{figure*}
\begin{figure*}
\centerline{\includegraphics[width=0.6\linewidth]{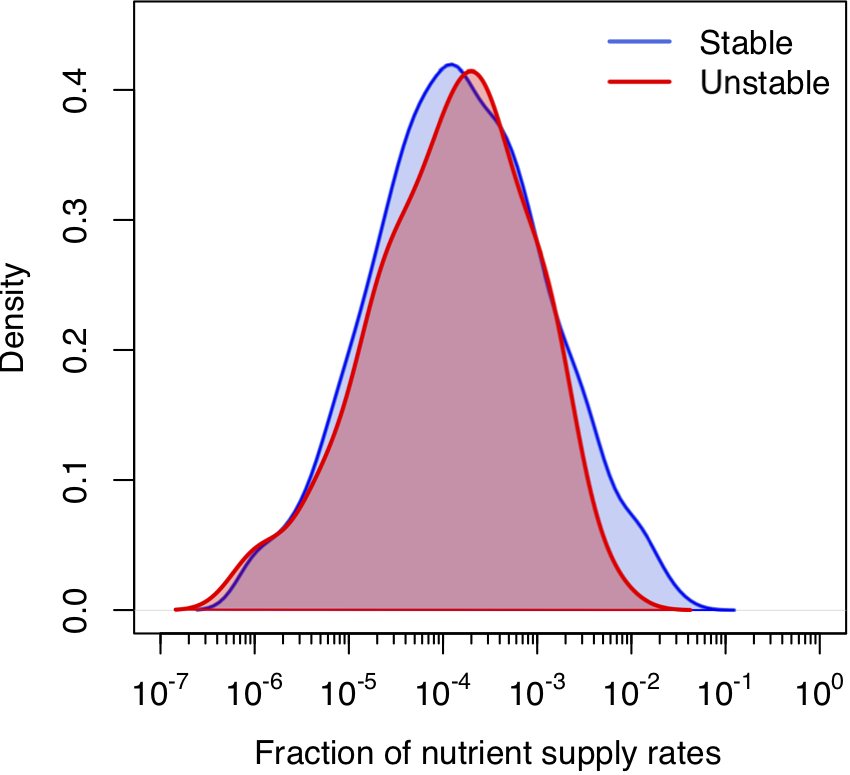}}
\caption{{\bf Distribution of volumes for stable and unstable states.} 
Log-normal distribution of feasible volumes of 
1211 uninvadable states in $6C\times6N\times36S$ version of our model.
Red line is used for 1058 dynamically stable states and blue line for 153 dynamically unstable ones. Distributions are normalized to the total number of states of each type.
The natural logarithm of the volume has mean 
$\mu=-8.87 \pm 0.06$ and standard deviation $\sigma=2.08 \pm 0.04$. 
There is no significant difference between distributions of volumes of 
stable and unstable uninvadable states (two-sample 
Kolmogorov-Smirnov test: p-value$=0.94$).
}
\label{figs1}
\end{figure*}
\begin{figure*}
\centerline{\includegraphics[width=0.99\linewidth]{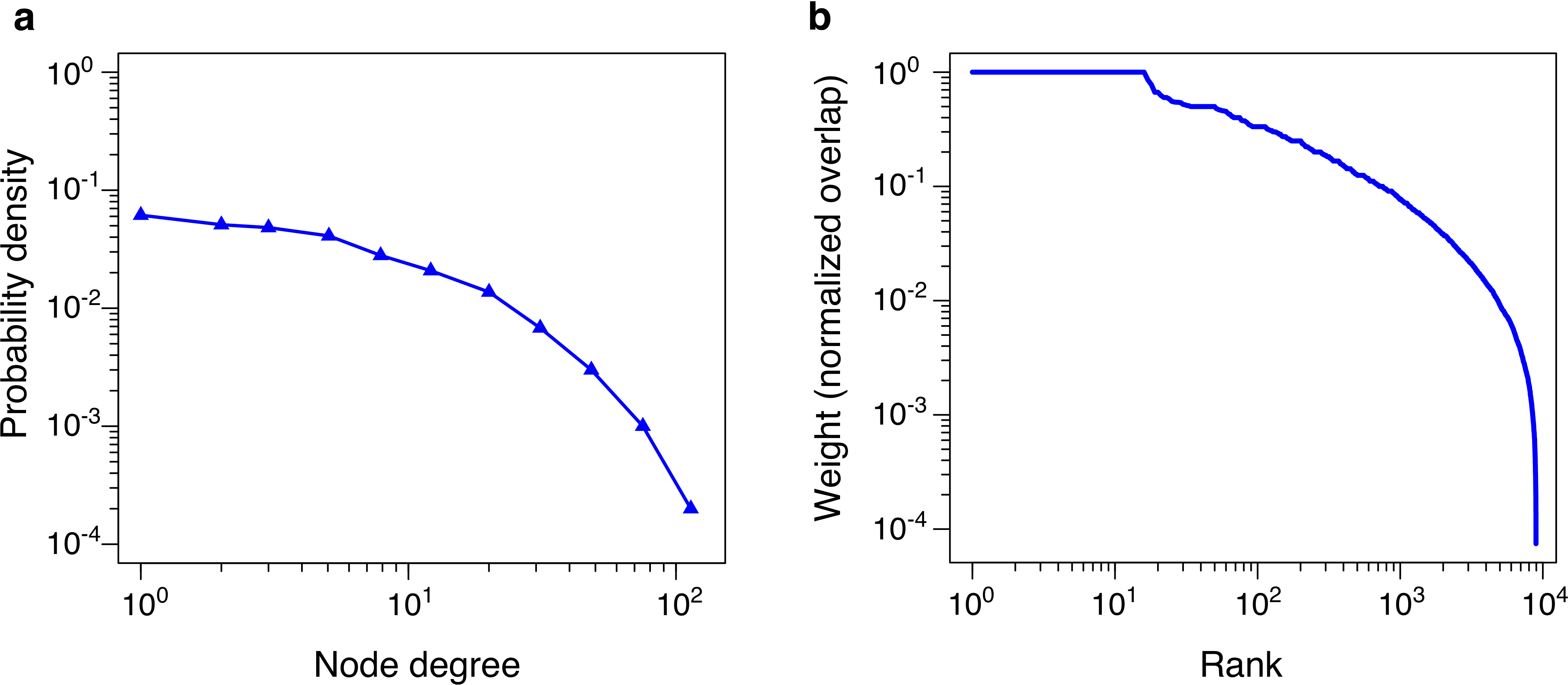}}
\caption{{\bf Statistics of pairwise bistability network for $6C\times6N\times36S$ example.} 
(\textbf{a}) Degree distribution of the network in Fig. \ref{fig2}F. 
(\textbf{a}) Rank-ordered distribution of weights of network edges from Fig. \ref{fig2}F. The weights of the network in Fig. \ref{fig2}F are given by normalized overlaps between states. 
}
\label{figs2}
\end{figure*}
\begin{figure*}
\centerline{\includegraphics[width=0.5\linewidth]{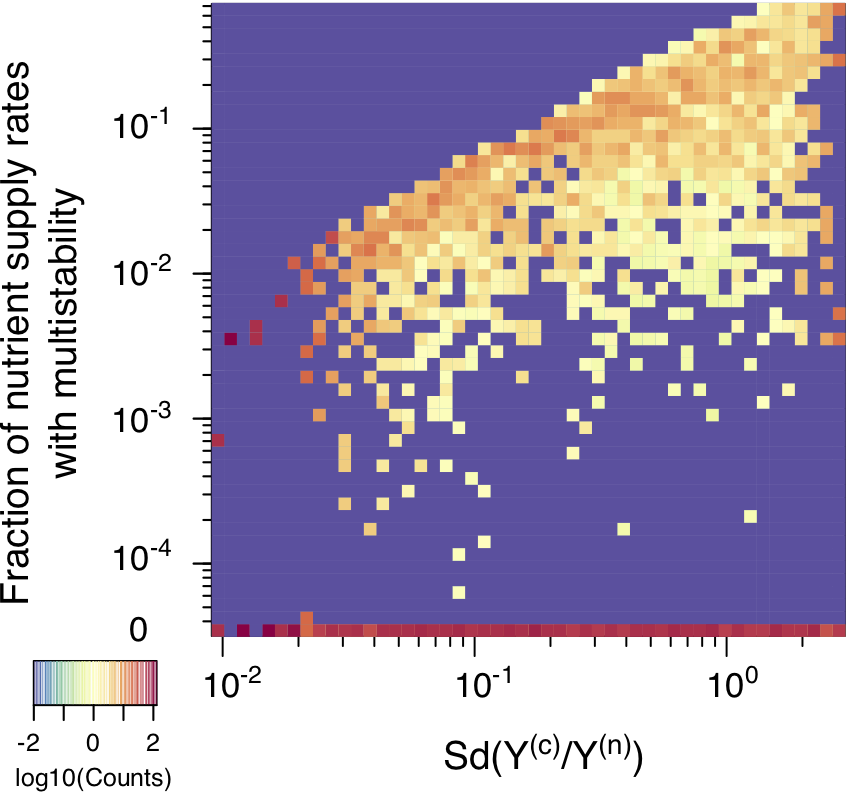}}
\caption{{\bf Statistics of multistability for different combinations of yields.} 
\textcolor{black}{Heatmap of the fraction of nutrient supply rate combinations that permit multistability for the $2C\times2N\times4S$ model with diverse combinations of microbial growth yields (4000 model variants in total) with diverse values of microbial growth yields . The color scale represents log10 of normalized counts of examples for a given interval of standard deviation of yields (X-axis) and the fraction of nutrient supply rates with multistability (Y-axis). 
The bottom row ($0$) corresponds to 2069 yields combinations where no multistability was observed. 
}
}
\label{figs3}
\end{figure*}
\begin{figure*}
\centerline{\includegraphics[width=\linewidth]{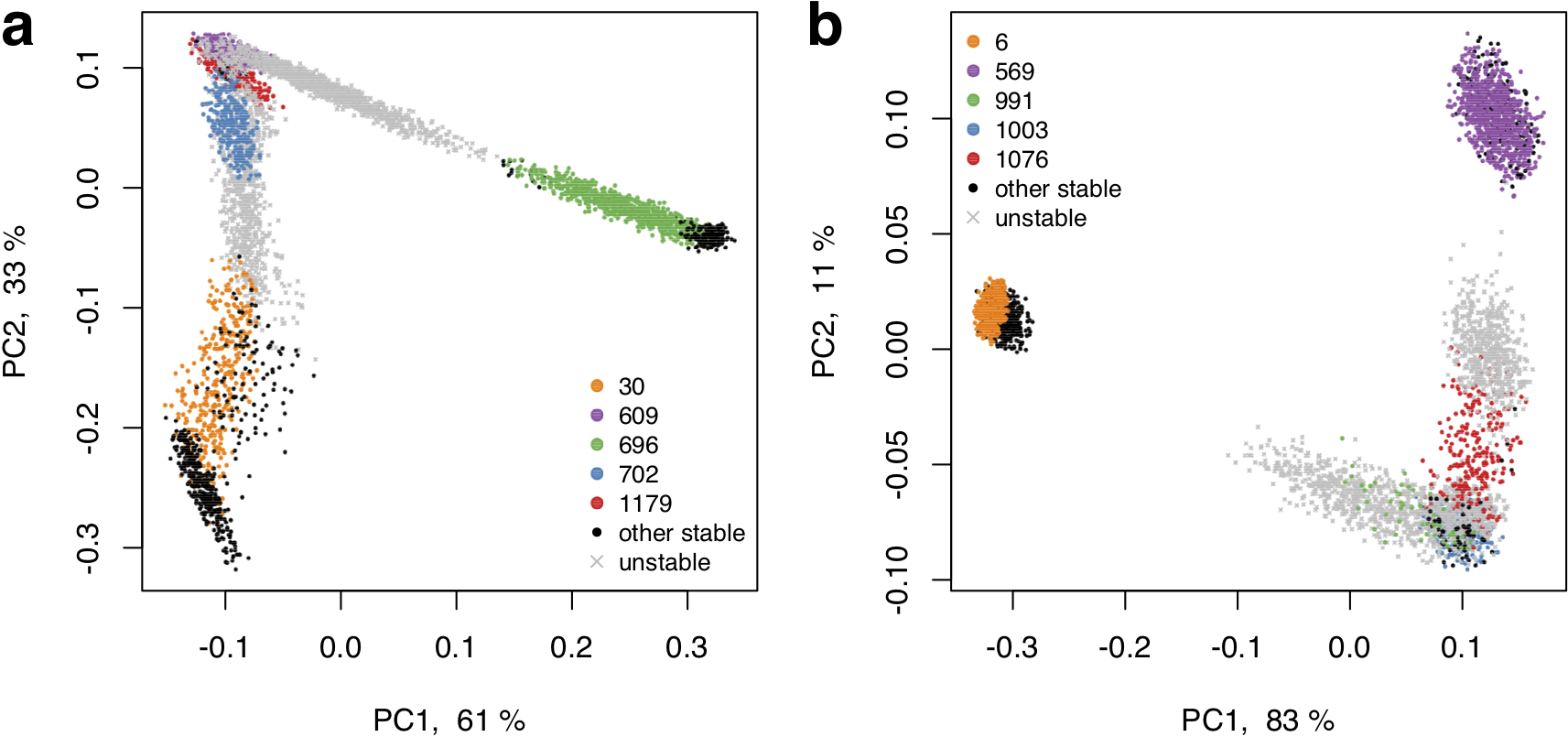}}
\caption{
{\bf The PCA plot of fractional
microbial abundances}. These abundances,  
normalized to 1, were obtained in our simulations for supply rates in the local vicinity of a multistable point
where $V=5$ stable states coexist. 
Panels (a) and (b) represent the two remaining 
multistable points in addition to the multistable 
point shown in Fig. \ref{fig3}C. Axes show the percentage of the variance explained by each principle component. Each point show 
the first (x-axis) and the second (y-axis) principal coordinates of microbial abundances in an 
uninvadable state feasible for a given set of nutrient supply rates. The supply rates were chosen to be close ($\pm$10\%) 
to the initial multistable point. Colored circles mark the original five stable states, 
black circles - other stable states which became feasible for
nearby supply rates, and grey crosses - dynamically unstable states feasible in this influx region. 
Note a quasi-1D manifold along which the points of all colors are aligned and 
the alternating order of points corresponding to stable and unstable states. 
}
\label{figs4}
\end{figure*}

3) Consider a steady state in which all nitrogen sources
are occupied. In our network representation it corresponds 
to every nitrogen source sending an outgoing link to some 
carbon source. Let $L_i$ be the number of microbes using 
the carbon source $i$ in a non-limiting fashion (the in-degree
of these outgoing links ending on $c_i$, see Fig. \ref{fig5}C). Then, obviously, 
$L = \sum L_i$ (note that some of the terms in this sum 
might be equal to zero). 

One can prove that if the state is uninvadable, then 
the matching given by all edges going from nitrogen sources 
to carbon sources must be stable in the Gale-Shapley sense.
To prove this, let's think of nitrogen sources as ``applicants'' and carbon 
sources as ``hospitals'' with their numbers of 
``vacancies'' given by $L_i$. Indeed, any unstable matching has at least one 
blocking pair $(n_j,c_i)$ such that:
\begin{itemize}
\item Condition 1. The nitrogen source (`applicant'') $n_j$ ``prefers'' the carbon source (``hospital'') $c_i$ to its 
currently assigned carbon source (the one used by the current microbe $B_{kj}$ limited 
$n_j$). This means that $\lambda^{(n)}_{ij} > \lambda^{(n)}_{kj}$. 
Thus the microbe $B_{ij}$ can grow on its nitrogen
source (provided that it can also grow on its carbon source).
\item Condition 2. The carbon source (``hospital'') $c_i$ ``prefers'' the nitrogen source (``applicant'') $n_j$ to 
at least one of $L_i$ of its currently assigned carbon sources  (the set of microbes using $c_i$ in a non-rate-limiting fashion). Thereby $\lambda^{(c)}_{ij}$ must be 
larger than the smallest $\lambda^{(c)}$ among these microbes.
According to the Exclusion Rule 2, this smallest $\lambda^{(c)}$
is still larger than $\lambda^{(c)}$ of the microbe limited by $c_i$ 
(if it exists). Thus the microbe $B_{ij}$ can also grow on its carbon 
source.
\end{itemize}
This proves that the microbe $B_{ij}$ corresponding to any blocking pair
can grow on both its carbon and its nitrogen sources, and thereby can 
successfully invade the steady state. This finishes the proof that any 
uninvadable state has to be a stable matching in the Gale-Shapley sense.

However, this does not prove that any stable matching 
corresponds to exactly one uninvadable state.
To prove this we first notice that, up to this point, our candidate 
uninvadable state contained only the 
nitrogen-limited species (See Fig. \ref{fig5} B) . We will now supplement it 
with carbon-limited species in such a way that 
1) added species do not violate the exclusion rule 2; 
2) added species render the state completely uninvadable. 
Let is introduce a new notation (applicable to our case in 
which all nitrogen sources are occupied). 
Let $\lambda^{(c)}_{\textrm{min}}(i)$ 
denote the smallest $\lambda^{(c)}$ among all 
species using $c_i$ in a non-rate-limiting fashion.
The Gale-Shapley theorem only guarantees the protection 
of our state from invasion by a species $(i,j)$ with 
$\lambda^{(c)}_{ij}$  
larger than $\lambda^{(c)}_{\textrm{min}}(i)$
(see the Condition 2 above). To ensure that 
our state is uninvadable by the rest of the 
species, one needs to add some carbon-limited species to 
this state. In order to do this in a systematic way, 
for each $c_i$ we compile the list of all species 
using this carbon source with $\lambda^{(c)}<\lambda^{(c)}_{\textrm{min}}(i)$. 
Each of these species is a potential invader.
Some species could be crossed off from the list 
of potential invaders because they cannot grow on their 
nitrogen source. 
These species have $\lambda^{(n)}$ below 
that of the (unique) species limited by their nitrogen source.
Among the species that remained on the list of 
invaders after this procedure, we select that with the largest 
$\lambda^{(c)}$ and add it to our steady state as 
a $C \to N$ directed edge, that is to say, as a 
carbon-limited species. This will prevent all 
other potential invaders on our list, since they 
have smaller $\lambda_{(c)}$ and thus, following the addition of our top carbon-limited species, they would 
no longer be able to grow based on their carbon source.
We will go over all $c_i$ and add such carbon-limited 
species if they are needed. The only scenario when 
such species is not needed if our list of potential invaders 
would turn up to be empty. In this case we will leave 
this carbon source unoccupied. See Fig. \ref{fig5} C
for the illustration of an uninvadabe state constructed by the above procedure in $6C\times6N\times36S$ model. Since for each carbon source the above algorithm selects the carbon-limited species (or selects to add no such species) in a unique fashion, there is a single uninvadable state for every stable matching in the Gale-Shapley sense. We are now in a position to predict and enumerate all uninvadable states in our model. 

\subsubsection*{Lower bound on the number of uninvadable states}
To count the number of partitions $(L_1, L_2, ..., L_L)$ such 
that $\sum L_i = L$, one can use a well known combinatorial 
method. According to this method, one introduces 
$L-1$ identical ``separators'' (marked with $|$) which are placed between 
$L$ identical objects (marked $\cdot$) separating them into $L$
(possibly empty) partitions. For example, for $L=4$ a partition 
$0,1,0,3$ would be denoted as $|\cdot||\cdot \cdot \cdot$. 
The combinatorial number of all possible arrangements of separators and 
objects is obviously ${{2L-1}\choose{L}}$.
For every such partition the Gale-Shapley theorem guarantees at least
one stable matching (that is, at least one
uninvadable steady state).
The lower bound on the number of uninvadable steady 
states has to be doubled to account for reversal of roles of 
carbons and nitrogens. There is a small possibility 
that we double counted one partition 
$(1, 1, ..., 1)$. Indeed, the unique uninvadable stable state corresponding to 
this partition could in principle be counted both when we start from nitrogen sources 
and when we start from carbon sources. This could happen only 
when the numbers of carbon
and nitrogen sources are equal to each other. More restrictively, 
this partition will be double-counted only if, when we started from C, 
all of the N-sources will send a link back to C, and these links all will end on different 
C-sources. The same has to be true if one starts with N-sources and 
at then sends links back to C. The steady state network in this case will consist of 
one or more loops covering all nutrients.
However, one can prove that, at least for the Gale-Shapley nitrogen-optimal state, 
the last carbon to be picked up would not need to send back a carbon-limited link.
Thus in our task of calculating the lower bound on the number of uninvadable states, 
we don't need to correct for the possibility of double-counting since at least 
one stable matching per partition (namely the Gale-Shapley) would not be double-counted.
Then we have  $N_{UIS} \geq 2{{2L-1}\choose{L}}={{2L}\choose{L}}$. 
The Sterling approximation for this expression is 
$2^{2L}/\sqrt{\pi L}$.  Thus the overall lower bound 
for the number of uninvadable stable states is given by 
\begin{equation}
  \label{eq:NUIS lower bound SI}
  N_{UIS}(L,L) \ge \cdot {{2L}\choose{L}} \simeq \frac{2^{2L}}{\sqrt{\pi L}} \qquad .
\end{equation}

More generally, the number of carbon sources, $K$, is not equal to the number 
of nitrogen sources, $M$. The resource type with a larger number will always 
have at least one resource left without input. Thus here one never 
needs to correct for double counting. Using the same 
reasoning as for $K=M=L$, the lower bound on the number of resources in this case 
is given by ${{K+M-1}\choose{K-1}}+{{K+M-1}\choose{M-1}}={{K+M}\choose{K}}$.
Here, the first term counts the uninvadable steady states in which all nitrogen sources 
are occupied and the partition divides $M$ edges sent by nitrogen sources 
among $K$ carbon sources, which requires $K-1$ ``dividers''. 
The second term counts the number of uninvadable steady states 
in which all carbon sources are occupied. Denoting 
the fraction of carbon resources among all resources as $p=K/(K+M)$ and 
using the Stirling approximation one gets 
\begin{eqnarray}
  \label{eq:NUIS hard case lower bound}
  &&N_{UIS}(K,M) \ge \cdot {{K+M}\choose{K}} \simeq \\ \nonumber 
&\simeq& \frac{\exp\left[(K+M)(-p \log p -(1-p) \log(1-p)\right]}{\sqrt{2\pi (K+M)p(1-p)}} \qquad .
\end{eqnarray}

%

In the case of multiple microbial species using the same
pairs of resources, our version of the
Gale-Shapley resident-oriented algorithm must be further
updated.  Let $M$ be the number of nitrogen sources,
and $K$ --- the number of carbon sources in the ecosystem,
$S$ the number of species in our pool, each requiring a pair of
resources to grow.  As now there may be more than
one microbe that uses a given pair of resources 
$c_i$ and $n_j$, we introduce
the notation $B_{ij}^{(r)}$ for the $r$th microbe
using the same pair of sources $c_i$ and $n_j$.  On average, 
each nitrogen (carbon) source has $S/K$ ($S/M$) 
microbes, which are capable of using it.  
As in the traditional Gale-Shapley algorithm, 
each nitrogen (carbon) source ranks all microbes 
capable of using it by their $\lambda^{(n)}$
($\lambda^{(c)}$).  

The way to identify all uninvadable stable states 
in this case is determined by a variant of the 
stable marriage problem (or rather the hospital/resident 
problem) in which every man (and every woman) 
may have more than one way to propose marriage to 
the same woman (man). In our model this corresponds 
to more than one microbe (a type of marriage) capable 
of growing on the same pair of carbon 
(corresponding to, say, men) 
and nitrogen (corresponding to women) sources. 
You may think of it as if each participant has 
several different ways to propose to the person of the opposite sex
(send flowers, take to a restaurant, etc). 
Each of these proposals is ranked by both parties 
independent of other ways. 
As far as we know, this variant has not been 
considered in the literature yet. However, all of the
results of the usual stable marriage (or hospital-resident)
problem remain unchanged. 

One can easily see that our lower bound (Eq. \ref{eq:NUIS hard case lower bound}) 
on the number of uninvadable states (equal to the number of 
stable marriages in all partitions) remains unchanged. Indeed, it is given by 
the number of partitions and hence depends only on $K$ and $M$ and not 
on $S$. However, for $S \gg K \cdot M$ one expects to have 
many more stable marriages 
for each partition. Thus the lower bound we have established 
is likely to severely underestimate 
the actual number of UIS in the ecosystem.

\subsection{Conditions of multistability in the $2C\times2N\times4S$ ecosystem}\label{textsupp:st4}
Below we consider the general case of a $2C\times2N\times4S$ ecosystem.
Let's assume that the selected set of $\lambda$ parameters
allow potentially unstable state in which 
all four species are present. These species form a single loop
in the network representation (like state S7 in our 2Cx2Nx4S example).
Without loss of generality we may assume that 
$\lambda-$parameters satisfy $\lambda_{11}^{(n)} > \lambda_{21}^{(n)}; \;
\lambda_{21}^{(c)} > \lambda_{22}^{(c)}; \; \lambda_{22}^{(n)} > \lambda_{12}^{(n)}; \;
\lambda_{12}^{(c)} > \lambda_{11}^{(c)}$. 
The 4-species loop is then formed by the links $C_1 \to N_1$,
$N_1 \to C_2$, $C_2 \to N_2$, and $N_2 \to C_2$
In all other cases we may rename C and N resources until 
the direction of the loop is as stated above.
%
%
The system also has two 
uninvadable steady states 
(A) in which two microbes ($N_1 \to C_1$ and $N_2 \to C_1$) 
are limited by their nitrogen sources and (B) in which two other 
microbes are limited by their carbon sources ($C_1 \to N_2$ and 
$C_2 \to N_1$). These two states have their regions of feasibility in the influx
space.  In order for these regions to 
overlap with each other, thereby resulting in bistability
within the overlapping region, the yield parameters of species 
and nutrient supply rates have to satisfy the following conditions.


For the state (A), the conservation laws read as
\begin{eqnarray}
    \dfrac{\Phi_1^{(c)}}{\delta} &=& c_1 + \dfrac{B_{11}}{Y_{11}^{(c)}} \nonumber \\
      \dfrac{\Phi_2^{(c)}}{\delta} &=& c_2 + \dfrac{B_{22}}{Y_{22}^{(c)}}  
      \label{eq:conservation A}\\
        \dfrac{\Phi_1^{(n)}}{\delta} &=& \dfrac{\delta}{\lambda_{11}^{(n)}} + \dfrac{B_{11}}{Y_{11}^{(n)}} \nonumber \\
        \dfrac{\Phi_2^{(n)}}{\delta} &=& \dfrac{\delta}{\lambda_{22}^{(n)}} + \dfrac{B_{22}}{Y_{22}^{(n)}} 
\nonumber        
\end{eqnarray}
The last two relations in (Eq. \ref{eq:conservation A}) define microbe concentrations as $B_{11} =
\dfrac{Y^{(n)}_{11}}{Y_{11}^{(c)}} \,
\Big(\,\dfrac{\Phi_1^{(n)}}{\delta}\, - \,
\dfrac{\delta}{\lambda_{11}^{(n)}}\Big)$, $B_{22} =
\dfrac{Y^{(n)}_{22}}{Y_{22}^{(c)}} \,
\Big(\,\dfrac{\Phi_2^{(n)}}{\delta}\, - \,
\dfrac{\delta}{\lambda_{22}^{(n)}}\Big)$.   Substituting
these expressions into the first two relations in  (Eq. \ref{eq:conservation A}) and invoking the requirements $c_1 >
\dfrac{\delta}{\lambda_{11}^{(c)}}, \; c_2 >
  \dfrac{\delta}{\lambda_{22}^{(c)}}$ (guaranteeing that
  neither of two carbons limits microbes' growth),
in the high flux limit  we obtain 
\begin{eqnarray}
        Y_{11}^{(n)} \Phi_1^{(n)} &<& Y_{11}^{(c)}
        \Phi_1^{(c)} \nonumber \\
        \label{eq:Y PHI conditions, state A} \\
         Y_{22}^{(n)} \Phi_2^{(n)} &<&  Y_{22}^{(c)}
        \Phi_2^{(c)} \nonumber 
\end{eqnarray}
The inequality conditions above are only natural given
that the microbes use up their nitrogen fluxes very
thoroughly in the state (A) while (at least in the 
high-flux limit) they not getting even close
to consuming all of their carbon supply rates .

The state (B) in the high flux limit will require
different conditions, though obtained in a perfectly
similar way:
\begin{eqnarray}
Y_{12}^{(n)} \Phi_2^{(n)} &>& Y_{12}^{(c)}
\Phi_1^{(c)} \nonumber \\
\label{eq:Y PHI conditions, state B}\\
Y_{21}^{(n)} \Phi_1^{(n)} &>&  Y_{21}^{(c)}
\Phi_2^{(c)} \nonumber 
\end{eqnarray}
Combining Eq. \ref{eq:Y PHI conditions, state A} and 
Eq. \ref{eq:Y PHI conditions, state B} one gets
\begin{eqnarray}
\dfrac{Y_{11}^{(c)}}{Y_{11}^{(n)}} \,
\Phi_1^{(c)} & >&  \Phi_1^{(n)} >  \;
\dfrac{Y_{21}^{(c)}}{Y_{21}^{(n)}} 
\, \Phi_2^{(c)} \nonumber \\
\label{eq:Y PHI conditions, A, B combined} \\
\dfrac{Y_{22}^{(c)}}{Y_{22}^{(n)}}
 \,  \Phi_2^{(c)} &>&  \Phi_2^{(n)} > \;
\dfrac{Y_{12}^{(c)}}{Y_{12}^{(n)}} \, \Phi_1^{(c)}
\nonumber
\end{eqnarray}
Hence, for the carbon fluxes ratio, one would have
\begin{equation}
      \label{eq:carbon flux ratio}
\dfrac{Y_{21}^{(c)} \, Y_{11}^{(n)}}{Y_{21}^{(n)}
 \,  Y_{11}^{(c)}} \, < \,
\dfrac{\Phi_1^{(c)}}{\Phi_2^{(c)}} \, < \,
\dfrac{Y_{22}^{(c)} \, Y_{12}^{(n)}}{Y_{22}^{(n)} \,
  Y_{12}^{(c)}}
\end{equation}
which implies that for multistability to be possible at least for some ration of fluxes the yields have to satisfy the following inequality:
\begin{equation}
      \label{eq:general bistability condition}
\dfrac{Y_{21}^{(c)} \, Y_{11}^{(n)}
 \,  Y_{22}^{(n)} \,
  Y_{12}^{(c)}}{Y_{21}^{(n)} \, Y_{11}^{(c)}
 \,  Y_{22}^{(c)} \,
  Y_{12}^{(n)}}\, < \, 1
\end{equation}
Note that the Eq. \ref{eq:general bistability condition} is both
necessary and sufficient for bistability between 
(A) and (B) for some set of supply rates.  Indeed, if Eq. \ref{eq:general bistability condition}
is satisfied, $\Phi_{2}^{c}$ can be chosen arbitrarily
(the only thing one would have to mind here is the 
high-flux limit requirements), then $\Phi_{1}^{c}$ should be chosen in accordance
with Eq. \ref{eq:carbon flux ratio}, and any nitrogen fluxes
satisfying Eq. \ref{eq:Y PHI conditions, A, B combined}.
All the procedures are legitimate whenever Eq. \ref{eq:general bistability condition} holds. 
Once chosen in the way described above, the point 
$(\Phi_{1}^{(c)}, \, \Phi_{2}^{(c)}, \, \Phi_{1}^{(n)}, \,
\Phi_{2}^{(n)})$ of the influx space will make both 
(A) and (B) steady states feasible.

In a more general case of an arbitrary number of nutrients of each type, the conditions allowing for bistability or even multistability can be 
expressed by simple inequalities connecting yields and fluxes in  combinations dictated by network topology of potentially bistable states
in a very similar way to the simple case 
presented above. Thus the solution to the 
puzzle of why roughly half of all possible yield combinations has no multistability whatsoever becomes intuitively clear.  
Indeed, these yields and
fluxes must come in ``dimensionless'' combinations so that any inequality can be written as a function of 
only C:N stoichiometry 
$S^{(C:N)}_{ij}=\dfrac{Y_{ij}^{(n)}}{Y_{ij}^{(c)}}$ 
for all microbial species present in any of the set of potentially multistable states. 
Note that  {\bfseries should nitrogen and
carbon yields exchange places for each of these microbes, the key inequality similar to 
Eq. \ref{eq:general bistability condition}
would be reversed}, thus prohibiting multistability
where it was permitted and vice versa.  

In the yield space, the proposed swap of carbon and nitrogen yields of all species $Y^{(c)} \to Y^{(n)}, \; Y^{(n)} \to Y^{(c)}$ is a volume preserving transformation.  This means that, for each set of potentially multistable states, the fraction of the yield space favouring multistability is exactly the same as the fraction prohibiting multistability.
In a possible (albeit unlikely) scenario when the
suggested permutation affects not only the
potential multistability in question, but also some other multistable conditions, ``turning off'' one multistability might in some cases ``turn on'' others.  This is why the ultimate empirical probability of multistability for a given combination of species' yields might somewhat deviate from $1/2$. 
\section{Supplementary tables}
\begin{table}[ht]
\caption{\label{tab:Lambdas_L2} $\lambda^{c}_{i,j}$, $\lambda^{n}_{i,j}$ values of the 4 species for the 2Cx2Nx4S model. 
}
\begin{tabular}{|P{0.8cm}|P{0.8cm}|P{0.8cm}||P{0.8cm}|P{0.8cm}|}
  \hline
  & \multicolumn{2}{c||}{$\lambda^{c}_{i,j}$} & \multicolumn{2}{c|}{$\lambda^{n}_{i,j}$}\\
  \hline
   & $n_1$ & $n_2$ & $n_1$ & $n_2$ \\
  \hline
  $c_1$ & 41 & 35 & 16 & 50 \\
  \hline
  $c_2$ & 52 & 56 & 27 & 44 \\
  \hline
\end{tabular}
\end{table}

\begin{table}[ht]
\caption{\label{tab:Yields_L2} Values of carbon and nitrogen Yields of the 4 species for the 2Cx2Nx4S model. 
}
\begin{tabular}{|P{0.8cm}|P{0.8cm}|P{0.8cm}||P{0.8cm}|P{0.8cm}|}
  \hline
   & \multicolumn{2}{c||}{$Y^{c}_{i,j}$} & \multicolumn{2}{c|}{$Y^{n}_{i,j}$}\\
   \hline
   & $n_1$ & $n_2$ & $n_1$ & $n_2$\\
  \hline
  $c_1$ & 0.37 & 0.64 & 0.27 & 0.10 \\\hline
  $c_2$ & 0.47 & 0.14 & 0.22 & 0.59 \\\hline
\end{tabular}
\end{table}
\begin{table}[ht]
\caption{\label{tab:Lambdas_c_L6} $\lambda^{(c)}_{(i,j)}$ values of the 36 species for the 6Cx6Nx36S model.
}
\begin{tabular}{|P{0.8cm}|P{0.8cm}|P{0.8cm}|P{0.8cm}|P{0.8cm}|P{0.8cm}|P{0.8cm}|}
  \hline
   & $N_1$ & $N_2$ & $N_3$ & $N_4$ & $N_5$ & $N_6$ \\\hline
   $C_1$ & 47.4 & 78.1 & 93.7 & 68.9 & 75.0 & 44.5 \\\hline
   $C_2$ & 89.6 & 68.6 & 33.6 & 77.8 & 16.5 & 90.8 \\\hline
   $C_3$ & 56.5 & 32.2 & 86.2 & 13.1 & 71.1 & 15.5 \\\hline
   $C_4$ & 53.0 & 94.1 & 38.7 & 10.7 & 34.0 & 34.9 \\\hline
   $C_5$ & 25.0 & 49.3 & 76.3 & 18.2 & 54.5 & 51.8 \\\hline
   $C_6$ & 47.1 & 91.9 & 57.7 & 63.0 & 92.2 & 90.0 \\\hline
\end{tabular}

\end{table}
\begin{table}[ht]
\caption{\label{tab:Lambdas_n_L6} $\lambda^{(n)}_{(i,j)}$ values of the 36 species for the 6Cx6Nx36S model. 
}
\begin{tabular}{|P{0.8cm}|P{0.8cm}|P{0.8cm}|P{0.8cm}|P{0.8cm}|P{0.8cm}|P{0.8cm}|}
  \hline
   & $N_1$ & $N_2$ & $N_3$ & $N_4$ & $N_5$ & $N_6$ \\\hline
   $C_1$ & 18.3 & 57.7 & 44.5 & 16.0 & 70.4 & 66.8 \\\hline
   $C_2$ & 56.7 & 31.4 & 78.6 & 91.8 & 34.5 & 34.7 \\\hline
   $C_3$ & 42.3 & 53.8 & 84.8 & 99.2 & 79.0 & 44.6 \\\hline
   $C_4$ & 95.3 & 91.4 & 73.1 & 42.9 & 98.8 & 66.7 \\\hline
   $C_5$ & 76.2 & 98.4 & 31.0 & 55.4 & 14.5 & 57.4 \\\hline
   $C_6$ & 37.6 & 79.3 & 58.4 & 71.8 & 26.0 & 84.5 \\\hline
\end{tabular}
\end{table}
\begin{table}[ht]
\caption{\label{tab:Yields_c_L6} $Y^{(c)}_{(i,j)}$ values of the 36 species for the 6Cx6Nx36S model. 
}
\begin{tabular}{|P{0.8cm}|P{0.8cm}|P{0.8cm}|P{0.8cm}|P{0.8cm}|P{0.8cm}|P{0.8cm}|}
  \hline
   & $N_1$ & $N_2$ & $N_3$ & $N_4$ & $N_5$ & $N_6$ \\\hline
   $C_1$ & 0.72 & 0.59 & 0.15 & 0.11 & 0.13 & 0.75 \\\hline
   $C_2$ & 0.29 & 0.72 & 0.79 & 0.39 & 0.15 & 0.16 \\\hline
   $C_3$ & 0.76 & 0.61 & 0.40 & 0.86 & 0.60 & 0.63 \\\hline
   $C_4$ & 0.87 & 0.68 & 0.64 & 0.27 & 0.80 & 0.51 \\\hline
   $C_5$ & 0.88 & 0.13 & 0.31 & 0.36 & 0.11 & 0.48 \\\hline
   $C_6$ & 0.46 & 0.34 & 0.38 & 0.83 & 0.70 & 0.86 \\\hline
\end{tabular}
\end{table}
\begin{table}[ht]
\caption{\label{tab:Yields_n_L6} $Y^{(n)}_{(i,j)}$ values of the 36 species for the 6Cx6Nx36S model. 
}
\begin{tabular}{|P{0.8cm}|P{0.8cm}|P{0.8cm}|P{0.8cm}|P{0.8cm}|P{0.8cm}|P{0.8cm}|}
  \hline
   & $N_1$ & $N_2$ & $N_3$ & $N_4$ & $N_5$ & $N_6$  \\\hline
   $C_1$ & 0.10 & 0.30 & 0.67 & 0.36 & 0.32 & 0.66 \\\hline
   $C_2$ & 0.83 & 0.30 & 0.47 & 0.30 & 0.40 & 0.58 \\\hline
   $C_3$ & 0.72 & 0.15 & 0.65 & 0.54 & 0.21 & 0.18 \\\hline
   $C_4$ & 0.30 & 0.22 & 0.84 & 0.64 & 0.29 & 0.56 \\\hline
   $C_5$ & 0.55 & 0.16 & 0.77 & 0.42 & 0.22 & 0.25 \\\hline
   $C_6$ & 0.49 & 0.67 & 0.89 & 0.80 & 0.50 & 0.19 \\\hline
\end{tabular}
\end{table}

%

\end{document}